\documentclass[aps,prb,twocolumn,groupedaddress,showpacs,showkeys]{revtex4}
\usepackage{graphics,graphicx}
\usepackage{color}
\usepackage{amssymb,amsmath}
\usepackage{bookmath}
\usepackage{bm}

\begin{document}
%~~~~~~~~~~~~~~~~~~~~~~~~~~~~~~~~~~~~~~~~~~~~~~~~~
\title{Unconventional superconductors under rotating magnetic field I: \\
density of states and specific heat}
\author{A. Vorontsov}
%\email[contact:]{anton@phys.lsu.edu}
\author{I. Vekhter}
\affiliation{Department of Physics and Astronomy,
     Louisiana State University, Baton Rouge, Louisiana, 70803, USA}
\date{\today}

\begin{abstract}
We develop a fully microscopic theory for the calculations of the
angle-dependent properties of unconventional superconductors under
a rotated magnetic field. We employ the quasiclassical Eilenberger
equations, and use a variation of the Brandt-Pesch-Tewordt (BPT)
method to obtain a closed form solution for the Green's function.
The equations are solved self-consistently for
quasi-two-dimensional $d_{x^2-y^2}$ ($d_{xy}$) superconductors
with the field rotated in the basal plane. The solution is used to
determine the density of states and the specific heat. We find
that applying the field along the gap nodes may result in minima
or maxima in the angle-dependent specific heat, depending on the
location in the $T$-$H$ plane. This variation is attributed to the
scattering of the quasiparticles on vortices, which depends on
both the field and the quasiparticle energy, and is beyond the
reach of the semiclassical approximation. We investigate the
anisotropy across the $T$-$H$ phase diagram, and compare our
results with the experiments on heavy fermion CeCoIn$_5$.
\end{abstract}
\maketitle
%~~~~~~~~~~~~~~~~~~~~~~~~~~~~~~~~~~~~~~~~~~~~~~~~~~~~~~~~~~~~~~~~~~~~~~~~~~~~~
%~~~~~~~~~~~~~~~~~~~~~~~~~~~~~~~~~~~~~~~~~~~~~~~~~~~~~~~~~~~~~~~~~~~~~~~~~~~~~

%~~~~~~~~~~~~~~~~~~~~~~~~~~~~~~~~~~~~~~~~~~~~~~~~~~~~~~~~~~~~~~~~~~~~~~~~~
\section{Introduction}
%~~~~~~~~~~~~~~~~~~~~~~~~~~~~~~~~~~~~~~~~~~~~~~~~~~~~~~~~~~~~~~~~~~~~~~~~~

In this paper and its companion~\cite{AVorontsov:II}, hereafter
referred to as II, we present a general theoretical approach for
investigation of thermal and transport properties of
superconductors in magnetic field, and use it to determine the
behavior of the density of states, specific heat, and thermal
conductivity in the vortex state of unconventional
superconductors. Our more specific goal here is to provide
connection between theory and recent experiments measuring the
properties of such superconductors under a rotating magnetic
field, to explain the existing data, and to guide future
experimental studies. We focus on these experiments as they hold
exceptional promise for helping determine the structure of the
superconducting energy gap.

We consider unconventional superconductors, for which in the
ordered state both the gauge symmetry and the spatial point group
symmetry are broken~\cite{SigristUeda}. Then the gap in the single
particle spectrum, $|\Delta(\hvp)|$, is momentum dependent. We
focus on anisotropic pairing states with zeroes, or nodes, of the
superconducting gap for some directions on the Fermi surface (FS).

The single particle energy spectrum of a superconductor is
$E(\hvp)=\pm\sqrt{\xi^2(\hvp)+|\Delta(\hvp)|^2}$, where
$\xi(\hvp)$ is the band energy in the normal state with respect to
the Fermi level. Consequently, the gap nodes, $|\Delta(\hvp)|=0$,
are the loci of the low energy quasiparticles, and the number of
quasiparticles excited by temperature or other perturbations
depends on the topology of the nodal regions. Experimental probes
that predominantly couple to unpaired electrons, for example the
heat capacity or (for pairing in the singlet channel)
magnetization, are commonly used to show the existence of the gap
nodes. The nodal behavior is manifested by $T^n$ power laws, with
the exponent $n$ that depends on the structure of the gap
\cite{SigristUeda}.

Locating the nodes on the Fermi surface is a harder task. Since
usually only the phase of the gap, but not the gap amplitude,
$|\Delta(\hvp)|$, breaks the point group symmetry, transport
coefficients in the superconducting state retain the symmetry of
the normal metal above $T_c$. The phase of the order parameter can
be tested by surface measurements, but experimental determination
of the nodal directions in the bulk requires breaking of an
additional symmetry. One possible approach is to apply a magnetic
field, $\vH$, and rotate it with respect to the crystal lattice.
The effect of $\vH$ on the nodal quasiparticles depends on the
angle between the Fermi velocity at the nodes and the field, and
hence provides a directional probe of the nodal
properties~\cite{IVekhter:1999R}.

At the simplest level, screening of the field and the resulting
flow of the Cooper pairs, either in the Meissner or in the vortex
state, locally shifts the energy required to create an unpaired
quasiparticle relative to the condensate (Doppler shift)
\cite{SKYip:1992,GVolovik:1993,DXu:1995}. Our focus here is on the
vortex state, where the supercurrents are in the plane normal to
the applied field, and hence only the quasiparticles moving in the
same plane are significantly affected. Applying the field at
different angles with respect to the nodes  preferentially excites
quasiparticles at different locations at the Fermi surface, and
leads to features in the density of states (as a function of the
field direction) \cite{IVekhter:1999R}. This, in turn, produces
oscillations in the measurable thermodynamic and transport
quantities, which can be used to  investigate the nodal structure
of unconventional superconductors.

Such investigations have been carried out experimentally in a wide
variety of systems. Due to higher precision of transport
measurements, more data exist on the thermal conductivity
anisotropy under rotated field. The anisotropy was reported in
high-temperature superconductors \cite{FYu:1995,HAubin:1997},
heavy fermion UPd$_2$Al$_3$ \cite{TWatanabe:UPd2Al3}, CeCoIn$_5$
\cite{KIzawa:CeCoIn5}, PrOs$_4$Sb$_{12}$ \cite{KIzawa:PrOsSb},
organic $\kappa$-(BEDT-TTF)$_2$Cu(NCS)$_2$ \cite{KIzawa:BEDT}, and
borocarbide (Y,Lu)Ni$_2$B$_2$C \cite{KIzawa:YNiBCkappa}, see
Ref.~\onlinecite{YMatsuda:2006} for review. The heat capacity
measurements are more challenging, and were carried out in the
borocarbides\cite{TPark:2003,TPark:2004}, and CeCoIn$_5$
\cite{HAoki:2004}. While the experiments provided strong
indications for particular symmetries of the superconducting gap
in these materials, they did not lead to a general consensus. The
main reason for that has been lack of reliable theoretical
analysis of thermal and transport properties in the vortex state.

Historically, there was a schism between theoretical studies of
the properties of $s$-wave type-II superconductors at low fields,
where the single particle states are localized in the vortex
cores, and the investigations near the upper critical field,
$H_{c2}$, where vortices nearly overlap and the quasiparticles
exist everywhere in space. The distinction between the two regimes
is not so clear cut in unconventional superconductors, since it is
the extended near-nodal states that control the electronic
properties both at high and at low fields \cite{GVolovik:1993}.
Often it is hoped that a single theoretical approach may provide
results valid over a wide temperature and field range in nodal
superconductors.

In part because early experiments on the vortex state of
unconventional superconductors focused on the high-T$_c$ cuprates
\cite{FYu:1995,HAubin:1997}, theoretical work has long been rooted
in the low field analysis. The Doppler shift approximation was
used to predict and analyze the behavior of the specific heat
\cite{GVolovik:1993,CKubert:1998SSC,IVekhter:1999R} and the
thermal conductivity
\cite{CKubert:1998,MFranz:1999,IVekhter:2000anis,PThalmeier:2002}
under an applied magnetic field. The method is semiclassical in
that it considers the energy shift of the nodal quasiparticles
with momentum $\hvp$ at a point $\vR$. Consequently, it is only
valid at low fields, $H\ll H_{c2}$, when the vortices are far
apart, and the supervelocity varies slowly on the scale of the
coherence length. Moreover, most such calculations account only
for quasiparticles near the nodes, and therefore are restricted to
energies small compared to the maximal superconducting gap, and
hence to temperatures $T\ll T_c$. In addition, the energy shift
leaves the quasiparticle lifetime infinite in the absence of
impurities, and therefore the method does not account for the
scattering of the electrons on vortices. While some attempts to
remedy the situation exist \cite{FYu:1995,MFranz:1999,WKim:2003},
no consistent description emerged.

Recent experiments cover heavy fermion and other low temperature
superconductors, and generally include the regime $T\lesssim T_c$
and $H\lesssim H_{c2}$. Consequently, there has been significant
interest in developing alternatives to the low field Doppler shift
approach. The goal is to treat transport and thermodynamics on
equal footing, to be able to describe the electronic properties
over a wide range of fields and temperatures, and to include the
effects of scattering on vortices. Fully numerical solution of the
microscopic Bogoliubov-de Gennes equations  have been employed for
computing the density of states (see, for example,
Ref.~\onlinecite{MUdagawa:2004}), but are not naturally suited for
computing correlation functions and transport properties.
Calculation of the Green's function in the superconducting vortex
state is difficult due to appearance of additional phase factors
from the applied field. Moreover, transport calculations need to
include the vertex corrections, since the characteristic
intervortex distance is large compared to lattice spacing, hence
the scattering on the vortices corresponds to small momentum
transfer, and the forward scattering is important.

Here we use the microscopic approach in conjunction with a variant
of the approximation originally due to Brandt, Pesch and Tewordt
(BPT) \cite{BPT:1967} that replaces the normal electron part of
the matrix Green's function by its spatial average over a unit
cell of the vortex lattice. While originally developed for
$s$-wave superconductors, this approach has recently been
successfully and widely applied to unconventional systems (see
Sections \ref{sec:QCvort} and \ref{sec:GreenSol} for full
discussion and references), where it gave results that are
believed to be valid over a wide range of temperatures and fields
\cite{IVekhter:1999,TDahm:2002}.

We employ the approximation in the framework of the quasiclassical
method \cite{eil68,lar68}. Two main advantages of this approach
are: a) BPT approximation results in a closed-form solution for
the Green's function
\cite{WPesch:1975,AHoughton:1998,IVekhter:1999} enabling us to
enforce self-consistency for any field, temperature, and impurity
scattering, and  facilitating the subsequent calculations of
physical properties; b) quasiclassical equations are
transport-like, so that the difference between single particle and
transport lifetimes appears naturally, without the need to
evaluate vertex corrections. Consequently, we are able to compute
the density of states, specific heat, and the thermal conductivity
on equal footing, and provide a detailed comparison with
experiment.

In this we pay particular attention to the data on heavy fermion
CeCoIn$_5$, where the specific heat and the thermal conductivity
data were interpreted as giving contradictory results for the
shape of the superconducting gap. The anisotropic contribution to
the specific heat exhibited minima for the field along the [100]
directions, which led the authors to infer $d_{xy}$
gap symmetry~\cite{HAoki:2004}, while the (more
complicated) pattern in the thermal conductivity for the heat
current along the [100] direction under rotated field was
interpreted as consistent with the $d_{x^2-y^2}$ gap
\cite{KIzawa:CeCoIn5}. In a recent Letter \cite{AVorontsov:2006}
we suggested a resolution for the discrepancy, and provide the
detailed analysis here.

The remainder of the paper is organized as follows. In
Sec.~\ref{sec:QC} we briefly review the quasiclassical approach
and the BPT approximation to the vortex state.
Sec.~\ref{sec:Green} gives the derivation of the equilibrium
Green's function. Some of the more technical aspects of the
calculation are described in the appendices: Appendix~\ref{app:GL}
describes a useful choice of ladder operators that enable us to
efficiently solve the quasiclassical equations in the BPT
approach, and Appendix~\ref{app:F} shows how to find a closed form
solution for the Green's function.

Many of the salient features of our results are clear from a
simple and pedagogical example of a 2D $d$-wave superconductor
with a cylindrical Fermi surface considered in Sec.~\ref{sec:CYL}.
We discuss the influence of the field on the density of states in
the vortex state and present the results for the anisotropy of the
specific heat and heat conductivity for an arbitrary direction of
the applied magnetic field.

%%%%%%%%%%%%%%%%%%%%%%%%%%%%%%%figure
\begin{figure}[t]
\centerline{\includegraphics[height=5.5cm]{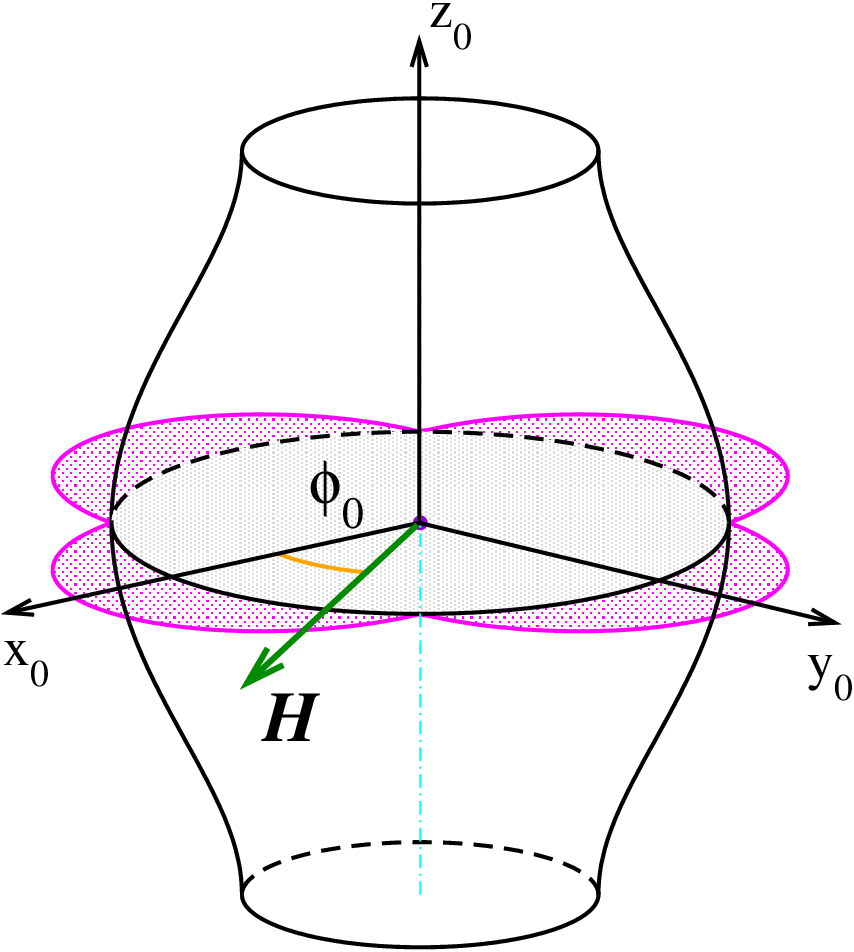}}
\caption{\label{fig:model} (Color online)
In this paper we present calculations
for a simple cylindrical Fermi surface
and a Fermi surface shown here.
The $d$-wave order parameter has lines of vertical nodes.
Our goal is a calculation of the thermodynamic properties, such
as specific heat and entropy, and their anisotropy under magnetic field
rotations, $\phi_0$, in the $ab$-plane.
}
\end{figure}
%%%%%%%%%%%%%%%%%%%%%%%%%%%%%%%figure
As one of our goals is the comparison of the results with the data
on layered CeCoIn$_5$, Sec.~\ref{sec:QCYL} is devoted to the fully
self-consistent calculations for more realistic quasi-cylindrical
Fermi surfaces, Fig.~\ref{fig:model}. The discussion of the
results, comparison with the data, and implications for future
experiments are contained in Sec.~\ref{sec:CON}. The companion
paper II uses these results to derive and  discuss the behavior of
the thermal conductivity.

We aimed to make the article useful to both theorists and
experimentalists. Sec.~\ref{sec:CYL} and Sec.~\ref{sec:CON} are
probably most useful for those readers who are interested only in
the overall physical picture and the behavior of the measured
properties; the figures in Sec.~\ref{sec:QCYL} show the main
differences between the self-consistent and non-self-consistent
calculations.

%~~~~~~~~~~~~~~~~~~~~~~~~~~~~~~~~~~~~~~~~~~~~~~~~~~~~~~~~~~~~~~~~~~~~~~~~~
\section{\label{sec:QC} Quasiclassical Approach}
%~~~~~~~~~~~~~~~~~~~~~~~~~~~~~~~~~~~~~~~~~~~~~~~~~~~~~~~~~~~~~~~~~~~~~~~~~

\subsection{Basic equations and formulation}

We begin by writing down the quasiclassical equations for a
singlet superconductor in magnetic field
\cite{eil68,lar68,ser83,ale85,AHoughton:1998,esc99} and
summarizing the details relevant for our discussion. The equations
are for the quasiclassical (low-energy $\vare$) Green's function,
which is a matrix in the Nambu (spin and particle-hole) space,
    \be \whg(\vR, \hvp; \vare) =
    \left( \begin{array}{cc}
    g & i\sigma_2 f \\
    i\sigma_2 \ul{f} & -g
    \end{array} \right) \,.
    \label{eq:dg}
    \ee
This matrix propagator has been integrated over the quasiparticle
band energy, and therefore depends only on the direction at the
Fermi surface, $\hvp$, and the center of mass coordinate, $\vR$.

We formulate our approach in terms of the real-energy retarded,
advanced, and Keldysh propagators. This is a natural path for the
self-consistent calculation of the quasiparticle spectrum, needed
for determination of thermodynamic properties such as entropy and
heat capacity. Moreover, the Keldysh technique is the most direct
route towards non-equilibrium calculations, required for the
transport properties such as thermal conductivity, which is
covered in the companion paper II. Consequently we establish a
unified approach to describe both the thermodynamics and transport
in the vortex state.

Retarded (R) and advanced (A) functions $\whg=\whg^{R,A}$ satisfy
(we take the electron charge $e<0$)
    \bea
    [(\vare + {e\over c} \vv_f(\hat{\vp}) \vA(\vR) )\, \widehat{\tau}_3
    - \whDelta(\vR, \hat{\vp}) - \whs_{imp}(\vR; \vare),
    \nonumber \\
    \whg(\vR, \hat{\vp}; \vare)] + i\vv_f(\hat{\vp}) \cdot \gradR \;
    \whg(\vR, \hat{\vp}; \vare) = 0 \,, \label{eq:eil}
    \eea
together with the normalization condition
    \be \whg^{R,A}(\vR,
    \hat{\vp}; \vare_m)^2 = -\pi^2 \widehat{1} \,. \label{eq:gnorm}
    \ee
Here $\vare$ is the real frequency, $\vv_f(\hat{\vp})$ is the
Fermi velocity at a point $\hat{\vp}$ on the FS. The magnetic
field is described by the vector potential $\vA(\vR)$, and the
self-energy $\whs$ is due to impurity scattering. The equations
for the retarded and the advanced functions differ in the
definition of the corresponding self-energies.

The mean field order parameter,
    \be \whDelta = \left(
    \begin{array}{cc}
    0 & i\sigma_2 \Delta \\
    i\sigma_2 \Delta^* & 0
    \end{array} \right) \,,
    \ee
is defined via the self-consistency equation involving the Keldysh
function $f^K$,
    \be
    \Delta(\vR,\hvp) = \int {d\vare \over 4\pi i} %\int {d\Omega_{\hvp'}\over4\pi}
    \int d\hvp'_\sm{FS} \; n_f(\hvp') \, V(\hvp, \hvp') \, f^K(\vR, \hvp'; \vare)
    \label{Eq:DeltaSC}\,,
    \ee
In equilibrium $f^K = (f^R-f^A) \tanh(\vare/2 T)$, and we obtain
the usual self-consistency equation computing the $\vare$-integral
in the upper (lower) half-plane for $f^R ~(f^A)$.

We wrote Eq.(\ref{Eq:DeltaSC}) for a general Fermi surface, and
therefore introduced the density of states (DOS) at a point $\hvp$
on the Fermi surface in the normal state, $N_f(\hvp)$. The net
density of states, ${\cal N}_f = \int d\hvp_\sm{FS} N_f(\hvp)$,
and we define $n_f(\hvp) = N_f(\hvp)/{\cal N}_f$. We absorbed the
net DOS, ${\cal N}_f$, into the definition of the pairing
potential, $V(\hvp, \hvp')$.

Since below we frequently perform the integrals over the Fermi
surface, we introduce a shorthand notation
    \be
    \langle \; \bullet \; \rangle_\sm{FS} = \int d\hvp_\sm{FS} \;
    n_f(\hvp) \;  \bullet \,,
    \ee
so that the the gap equation above can be rewritten as
    \be
    \Delta(\vR,\hvp) = \int {d\vare \over 4\pi i}\,  \langle \, V(\hvp,
    \hvp') \, f^K(\vR, \hvp'; \vare)\rangle_\sm{FS}
    \label{Eq:DeltaSC1}\,.
    \ee
All calculations below are for separable pairing,
    \be
    V(\hat{\vp},\hat{\vp}') = V_s \, \cY(\hat{\vp}) \, \cY(\hat{\vp}')
    \,,
    \ee
where $\cY(\hat{\vp})$ is the normalized basis function for the
particular angular momentum, $\langle
\cY(\hat{\vp})^2\rangle_{\sm{FS}} = 1$. For example, for
$d_{x^2-y^2}$ gap over a Fermi surface parameterized by angle
$\phi$, we have $\cY(\phi) = \sqrt 2 \cos 2\phi$. Hence the order
parameter is $\Delta(\vR,\hat{\vp}) = \Delta(\vR) \cY(\hat{\vp})$.

Finally, we include the isotropic impurity scattering via the
self-energy,
    \be \whs_{imp}(\vR; \vare) = \left( \begin{array}{cc}
    D+\Sigma & i\sigma_2 \Delta_{imp} \\
    i\sigma_2 \ul{\Delta}_{imp} & D-\Sigma
    \end{array} \right) = n_{imp} \hat{t}(\vR;\vare) \,.
    \label{Eq:sigma-imp}
    \ee
Here $n_{imp}$ is the impurity concentration, and, in the
self-consistent $t$-matrix approximation,
    \be \hat{t}(\vR; \vare)
    = u \hat{1} + u {\cal N}_f \langle \whg(\vR,\hvp;\vare) \rangle_\sm{FS}
    \; \hat{t}(\vR; \vare)\,,
    \label{Eq:t-matrix}
    \ee
where $u$ is the single impurity isotropic potential. Comparing
Eq.~(\ref{eq:eil}) and Eq.~(\ref{Eq:sigma-imp}) we see that
$\Sigma$ effectively renormalizes the energy $\vare$, while
$\Delta_{imp}$ accounts for the impurity scattering in the
off-diagonal channel. The term $D\,\widehat{1}$ drops out of
equations for the retarded and advanced Green's functions since
the unit matrix commutes with the Green's function in
Eq.~(\ref{eq:eil}). This term, however, generally appears in the
Keldysh part, and has a substantial effect on transport
properties.
\cite{PJHirschfeld:1988,lof04} %(Appendix~\ref{app:Keldysh}).
Below we parameterize the scattering by the ``bare'' scattering rate,
$\Gamma=n_{imp}/\pi {\cal N}_f$, and the phase shift $\delta_0$ of
the impurity scattering, $\tan \delta_0=\pi u {\cal N}_f$.

\begin{widetext}
In equilibrium we explicitly write Eqs.
(\ref{eq:eil})-(\ref{eq:gnorm}) as a system of equations,
    \begin{subequations}
    \begin{eqnarray}
    &&g^2 - f\,\ul{f} = -\pi^2 \,,
    \label{gnorm}
    \\
    &&i\vv_f(\hat{\vp}) \cdot \gradR \;
    g + \widetilde{\Delta}  \ul{f} - \widetilde{\ul{\Delta}}  f = 0 \,, %\nonumber
    \label{gtriv}
    \\
    &&\left[-2 i \tilde{\vare} + \vv_f (\hat{\vp}) \left( \gradR -   \frac{2ie}{c} \vA(\vR)
    \right)\right]f=2i\widetilde{\Delta} g\, ,
    \label{eqf}
    \\
    &&\left[-2 i \tilde{\vare} - \vv_f (\hat{\vp}) \left( \gradR +   \frac{2ie}{c} \vA(\vR)
    \right)\right] \ul{f} =2i \ul{\widetilde{\Delta}} g\,,
    \label{eqf1}
    \end{eqnarray}
    \label{Eq:basic}
    \end{subequations}
where $\tilde{\vare} = \vare - \Sigma$, $\widetilde{\Delta}=\Delta
+ \Delta_{imp}$, and $\widetilde{\ul{\Delta}}=\Delta^* + \ul{\Delta}_{imp}$.
\end{widetext}

\subsection{\label{sec:QCvort}
Vortex state ansatz and Brandt-Pesch-Tewordt approximation}

So far our discussion remained completely general. In the vortex
state of a superconductor, the order parameter and the field vary
in space, and the quasiclassical equations have to be solved
together with the self-consistency equations for the gap function,
and Maxwell's equation for the self-consistently determined
magnetic field and the vector potential. Finding a general {\em
non-uniform} solution of such a system is a daunting, or even
altogether impossible, task. Therefore we make several simplifying
assumptions and approximations that allow us to obtain a closed
form solution for the Green's function.

First, we assume the magnetic field to be uniform. This assumption
is valid for fields $H\gg H_{c1}$, where the typical intervortex
spacing (of the order of the magnetic length, $\Lambda = (\hbar
c/2|e|B)^{1/2}$) is much smaller than the penetration depth, the
diamagnetic magnetization due to the vortices is negligible
compared to the applied field, and the local field is close to the
applied external field, $B\approx H$. All the materials for which
the anisotropy measurements have been performed are extreme
type-II superconductors, where this assumption is valid over
essentially the entire field range below $H_{c2}$.

In writing the quasiclassical equations we only included the
orbital coupling to the magnetic field, assuming that it dominates
over the the paramagnetic (Zeeman) contribution. This is valid for
most superconductors of interest, and the detailed analysis of the
Zeeman splitting will be presented separately
~\cite{AVorontsov:Zeeman};the main conclusions of this paper
remain unaffected.

Second, we take an Abrikosov-like vortex lattice ansatz for the
spatial variation of the order parameter, which is a linear
superposition of the single-vortex solutions in the plane normal
to the field. We enforce the self-consistency condition, which
requires going beyond the simple form suggested by the linearized
Ginzburg-Landau equations. The details of this choice are given in
Sec.\ref{sec:Green} below.

In the vortex lattice state the quasiclassical equations generally
do not allow solution in a closed form. We therefore employ a
variant of the approximation originally due to Brandt, Pesch and
Tewordt (BPT) \cite{BPT:1967}. The method consists of replacing
the diagonal part of the Green's function by its spatial average,
while keeping the full spatial structure of the off-diagonal
terms. It was initially developed to describe superconductors near
the upper critical field, where the amplitude of the order
parameter is suppressed throughout the bulk, and the approximation
is nearly exact. This is confirmed by expanding the Green's
function in the the Fourier components of the reciprocal vortex
lattice, $g(\vR,\hvp;\vare)=\sum_\vK \; g(\vK,\hvp;\vare)
\exp(i\vK \vR)$. and noticing that $g^{R}(\vK)
\propto\exp(-\Lambda^2 {\vK}^2)$ so that the $\vK=0$ component is
exponentially dominant.\cite{BPT:1967} In situations where the
states inside vortex cores are not crucial for the analysis, such
as in extreme type-II ,\cite{EHBrandt:1976} or nodal
superconductors \cite{HWon:1996,IVekhter:1999} the method remains
valid essentially over the entire field range. Consequently the
BPT approach and its variations was extensively used to study
unconventional superconductors in the vortex
state.\cite{IVekhter:1999,HKusunose:2004,LTewordt:2005} One of the
advantages of the method that it reproduces correctly the $H=0$
BCS limit,\cite{AHoughton:1998} and therefore may be used to
interpolate over all fields. One, however, needs to be cautious in
computing the properties of impure systems: averaging over the
intervortex distance ($\sim\Lambda$) prior to averaging over
impurities is allowed only when $\Lambda/\ell \ll 1$, where $\ell$
is the mean free path, and hence the approach does break down at
very low fields, and only asymptotically approaches the zero field
result. We show the signatures of this breakdown in
Sec.~\ref{sec:QCYL}.

The use of the BPT approximation relaxes the constraints imposed
by the assumption of a perfectly periodic vortex arrangement.
Indeed, averaging over the unit cell of the vortex lattice is
somewhat akin to the coherent potential approximation in many body
physics, although with an important caveat that this is only done
for the normal part of the matrix Green's function. Consequently,
the results derived within this approach are also applicable to
moderately disordered vortex solids.

%\section{Single-particle Green's function and the density of states}
\section{Single-particle Green's function}
\label{sec:Green}

%\subsection{Solution for the Green's function}
\label{sec:GreenSol}

Hereafter we use $g$ to denote the spatially averaged electron
Green's function, $g \equiv g(\hat{\vp}; \vare) = \overline{g(\vR,
\hat{\vp}; \vare)}$. The approach we take here follows the
standard practice
\cite{WPesch:1975,AHoughton:1998,IVekhter:1999,HKusunose:2004} of
determining $g$ from the spatially averaged normalization
condition, Eq.(\ref{gnorm}),
    \be g^2 -
    \overline{\ul{f} \, f} = -\pi^2 \,.
    \ee
Here we defined the average over vortex lattice of a product as
    \be \overline{f_1\, f_2} = \int {d\vR \over V} \,
    f_1(\vR) f_2(\vR) \,.
    \ee
The anomalous components of the Green's function satisfy
Eqs.(\ref{eqf})-(\ref{eqf1}). Formally, the solution is obtained
by acting with the inverse of the differential operator in the
right hand side on the product $\widetilde\Delta \, g$ and
$\ul{\widetilde\Delta} \, g$ respectively. Upon replacement of $g$
by its average, the operator acts solely on the order parameter,
    \begin{subequations}
    \bea
    f(\vR, \hat{\vp}; \vare) = && 2 i g(\hat{\vp}; \vare)  \;
    \hat{O}_f \, \widetilde{\Delta}(\vR,\hat{\vp}; \vare)
    \\
    \ul{f}(\vR, \hat{\vp}; \vare) = && 2 i g(\hat{\vp}; \vare) \;
    \hat{O}_{\ul{f}}^* \, \widetilde{\ul\Delta}(\vR,\hat{\vp}; \vare)\, ,
    \eea
    \label{Eq:BPT}
    \end{subequations}
where
    \begin{subequations}
    \label{eq:O}
    \bea \hat{O}_f &=& [-2 i \tilde{\vare} +
    \vv_f(\hat{\vp})\,(\gradR -   i {2e\over c} \vA(\vR) )]^{-1} \,,
    \\
    \hat{O}_{\ul{f}}^* &=& [-2 i \tilde{\vare} - \vv_f(\hat{\vp})\,
    (\gradR +   i {2e\over c} \vA(\vR) )]^{-1} \,.
    \eea
    \end{subequations}

The strategy is to use a vortex lattice solution as an input,
compute the anomalous Green's functions $f$ and $\ul{f}$ in terms
of $g$ from Eq.(\ref{Eq:BPT}), determine $g$ from the
normalization condition, and then enforce the self-consistency on
$\Delta$ and the impurity self-energies. In principle, any
complete set of basis functions is suitable for expanding both
$\Delta(\bm R,\hat\vp)$ and $f(\bm R,\hat\vp)$. In practice, of
course, we are looking for an expansion that can be truncated
after very few terms, enabling efficient computation of the
functions. The Abrikosov lattice ansatz for $\Delta(\bm R)$ is a
superposition of the functions corresponding to the single vortex
solution of the Ginzburg-Landau equations, and therefore it is
natural to use these functions as our basis.

For an $s$-wave superconductor with an axisymmetric Fermi surface
(isotropic in the plane normal to the field), it is well known
that the vortex lattice is given by a superposition of the single
flux line solutions, the oscillator (Landau level, or LL)
functions, $\Phi_0(x-x_0)$, centered at different points in the
plane normal to the applied field \cite{Tinkham}
\begin{equation}
  \Delta(\bm R)=\sum_{k_y} C_{k_y} e^{ik_y y}
     \Phi_0\left( {x-\Lambda^2 k_y \over \Lambda} \right) \, .
    \label{AVL:s}
\end{equation}
Here the symmetry of the coefficients $C_{k_y}$ determines the
structure of the lattice. This form emerges from the solution of
the linearized Ginzburg-Landau (GL), and is also consistent with
the solution of the linearized, with $g=-i\pi$, quasiclassical
equations. Moreover, this form is valid down to low fields as the
admixture of the contributions from higher Landau levels, $\Phi_n$
with $n\neq 0$, to $\Delta(\bm R)$ remains
negligible.\cite{DLi:1999} Consequently, the set of oscillator
functions, $\Phi_n$, provides a convenient basis for the expansion
of anomalous functions $f$.  It is common to rewrite the operator
$\hat{O}$ via the bosonic creation and annihilation operators,
$a^\dag$ and $a$. \cite{AHoughton:1998} At the microscopic level,
inserting this ansatz for $\Delta(\bm R)$ into the quasiclassical
equations, Eqs.(\ref{Eq:BPT}), and enforcing the self-consistency
condition, yields the order parameter which only includes the
ground state oscillator functions, justifying use of
Eq.~(\ref{AVL:s}) \cite{AHoughton:1998}.

In unconventional superconductors the situation is more complex.
While the solution of the GL equations are still given by Eq.
(\ref{AVL:s}), this form is not a self-consistent solution of the
linearized microscopic equations: the momentum and the real space
dependence of the order parameter are coupled via the action of
the operator $\vv_f(\hat{\vp})\cdot\gradR$. Since the wave
functions for Landau levels form a complete set, they can still be
used as a basis for the expansion. The microscopic equations mix
different Landau levels, and the self-consistent solution for the
vortex state involves a linear combination of an infinite number
of $\Phi_n$ at each site \cite{ILukyanchuk:1987}. For the
axisymmetric case the spatial structure of $\Delta(\bm R)$ is
still close to that for the $s$-wave case, and the weight of the
higher Landau levels in the self-consistent solution decreases
rapidly with increasing $n$ \cite{ILukyanchuk:1987,HWon:1996}.
Hence in practice the series in $n$ is truncated either at $n=0$
(as for $s$-wave) or at the second non-vanishing term
\cite{HWon:1996,IVekhter:1999}. While this is often sufficient to
describe the salient features of the thermal and transport
coefficients, care should be taken in determining the anisotropies
of these coefficients under a rotated field: the anisotropy is
often of the order of a percent, and the structure of the vortex
lattice should therefore be determined to high accuracy as well.

The situation is even more complex for unconventional
superconductors with non-spherical Fermi surface, when the Fermi
velocity is anisotropic in the plane normal to the applied field.
Quasi-two dimensional systems with the field in the plane, such as
shown in Fig.~\ref{fig:model}, give one example of such
difficulties. Frequently in the microscopic theory the expansion
is still carried out in the LL functions using the operators for
the isotropic case. These functions  are now strongly mixed, and
hence (numerically intensive) inclusion of many LL is required
before the self-consistency is reached. Determining magnetization
in the vortex state, for example, was carried out with 6 LL
functions \cite{HAdachi:2005}.

This difficulty, however, is largely self-inflicted since, in
contrast to the isotropic case, the LL functions in the form used
in Ref.~\onlinecite{HAdachi:2005} are {\em not} the solutions to
the linearized GL equations. For an arbitrary Fermi surface the
coefficients of the Ginzburg-Landau expansion are anisotropic, and
the vortex lattice solution is given by the $n=0$ Landau Level in
the rescaled, according to the anisotropy,
coordinates.~\cite{Ketterson} We show in Appendix~\ref{app:GL}
that the proper rescaling is
    \be
    x' = x/\sqrt{S_f} , \qquad y' = y \sqrt{S_f} \,,
    \label{eq:xyscale}
    \ee
where $S_f$ is a measure of the anisotropy of the Fermi surface.
For a FS with rotational symmetry around the axis $z_0$, and for
the field at an angle $\theta_H$ to this axis,
    \be S_f = \sqrt{ \cos^2 \theta_H +
    {v_{0||}^2\over v_{0\perp}^2} \sin^2 \theta_H}  \,.
    \ee
Here  $v_{0\perp}^2 = 2 \langle \cY^2(\hvp) v^2_{\perp i}(p_z)
\rangle_\sm{FS}$ and $v_{0\parallel}^2 = 2 \langle \cY^2(\hvp)
v^2_\parallel(p_z) \rangle_\sm{FS}$, where $v_\parallel$ is the
projection of the Fermi velocity on the $z_0$ axis, and $v_{\perp
i}$ with $i=x_0,y_0$ is the projection on the axes in the plane
normal to $z_0$. For the field in the basal plane
$\theta_H=\pi/2$, and therefore $S_f=v_{0||}/ v_{0\perp}$.

The appropriate basis functions, which we use hereafter,
correspond to the oscillator states in the rescaled coordinates.
If we chose the direction of the field as the $z$-axis,
\begin{equation}
  \widetilde\Phi_n(x,k_y)=\Phi_n\left( {x-\Lambda^2
    \sqrt{S_f} k_y\over \Lambda \sqrt{S_f}} \right) \,.
    \label{Phi-n}
\end{equation}
For an $s$-wave superconductor the $n=0$ ansatz for $\Delta(\bm
R)$ satisfies microscopic equations, while for unconventional
order parameters different LLs are once again mixed. However, with
our choice of the basis functions this mixing is weak, enabling us
to truncate the expansion at three components. Consequently, we
use a generalized form of the vortex lattice $\Delta (\bm R,
\hvp)=\Delta(\bm R)\cY(\hvp)$, where
    \begin{subequations}
     \bea
     &&\Delta(\bm R)= \sum_n\Delta_n \braket{\vR}{n}
     \label{Eq:Dn}
     \\
     &&\braket{\vR}{n}=\sum_{k_y} C_{k_y}^{(n)} {e^{ik_y\sqrt{S_f} y}
    \over \sqrt[4]{S_f \Lambda^2}} \widetilde\Phi_n\left( x,k_y \right) \,.
    \label{Eq:Rn}
    \eea
    \label{Eq:VL}
    \end{subequations}
The normalizing factor in Eq.(\ref{Eq:Rn}) is introduced so that
the states $\braket{\vR}{n}$ are
orthonormal, i.e.
\begin{equation}
  \int \frac{d{\bm R}}{V} \braket{\vR}{n}
  [\braket{\vR}{n^\prime}]^*=\delta_{n,n^\prime}\, ,
\end{equation}
provided
\begin{equation}
  \sum_{k_y} | C_{k_y}^{(n)} |^2=1.
\end{equation}
Consequently $\Delta_n$ in Eq.(\ref{Eq:Dn}) has the meaning of the
amplitude of the appropriate component of the order parameter in
the LL expansion.

The ladder operators,
\begin{subequations}
    \bea
    a = {\Lambda\over \sqrt{2}} \left( -\grad_{x'} + i (\grad_{y'} +
    i{x'\over \Lambda^2}) \right) \,,
    \\
    a^\dag = {\Lambda\over \sqrt{2}} \left( \grad_{x'} + i (\grad_{y'}
    + i{x'\over \Lambda^2}) \right) \,,
    \eea
    \label{Eq:ladder}
\end{subequations}
obey the usual bosonic commutation relations, $[a, a^\dagger]=1$,
$[a,a]=[a^\dag,a^\dag]=0$, and connect the states $|n\rangle$ via
$a \ket{n} = \sqrt{n} \ket{n-1}$ and $a^\dag \ket{n} = \sqrt{n+1}
\ket{n+1}$.

To solve Eq.(\ref{Eq:BPT}) we rewrite the differential operators
$O_f$ and $O_{\ul f}$ via the ladder operators,
Eq.(\ref{Eq:ladder}), and  find
\begin{eqnarray}
  O_f&=&[-2 i \tilde{\vare} +
    \vv_f(\hat{\vp})\,(\gradR -   i {2e\over c} \vA(\vR) )]^{-1}
    \\
    &=&
    \left[-2 i \tilde{\vare} +
    \frac{1}{\sqrt 2
    \Lambda}\,\left( v_-(\hvp)a^\dagger-v_+(\hvp)a \right)\right]^{-1},
\end{eqnarray}
where
    \begin{equation}
      v_\pm=v_x(\hvp)/\sqrt{S_f}\pm i v_y(\hvp)\sqrt{S_f}
    \end{equation}
For convenience we introduce the  rescaled Fermi velocity
    \be
    \tilde{v}_f(\hat{\vp})_x = v_f(\hat{\vp})_x /\sqrt{S_f}  \quad ,\quad
    \tilde{v}_f(\hat{\vp})_y = v_f(\hat{\vp})_y \sqrt{S_f} \,,
    \ee
and its projection on the $xy$-plane (perpendicular to {\bf H}),
    \be
    |\tilde{v}_f^\perp (\hat{\vp})| =
    \sqrt{\tilde{v}_f(\hat{\vp})_x^2 + \tilde{v}_f(\hat{\vp})_y^2}\,,
    \ee
as well as the ``phase factors'',
    \be \tilde{v}_\pm (\hat{\vp})=
    \frac{\tilde{v}_f(\hat{\vp})_x \pm i \tilde{v}_f(\hat{\vp})_y}
      {|\tilde{v}_f^\perp|} \,.
    \ee

The off-diagonal parts of the matrix Green's function can be
expressed in terms of the normal component $g$, and written as a
series over the set $\braket{\vR}{m}$. The solution is based on
exponentiating the operator $O_f$ to explicitly evaluate the
result of its action on the order parameter
\cite{WPesch:1975,AHoughton:1998}, and is detailed in
Appendix~\ref{app:F}. We find
\begin{subequations}
    \be
    f(\vR, \hat{\vp}; \vare) = \sum_m f_m(\hvp, \vare) \braket{\vR}{m} \,,
    \ee
    \be
    f_m(\hvp, \vare) = ig \, \sum_n (-\tilde{v}_-(\hvp))^{m-n} \,
    \cD_{m,n}(\vare, |\hvp|) \widetilde{\Delta}_n(\hvp; \vare) \,,
    \ee
    \label{Eq:f-res}
\end{subequations}
where $\widetilde{\Delta}_n(\hvp; \vare) = \Delta_n(\hvp) +
\Delta_{imp,n}(\vare)$. The coefficients
    \be
    \cD_{m,n}(\vare, |\hvp|) = \sqrt{\pi}{2 \Lambda \over |\tilde{v}^\perp_f|}
    \sum_{j=0}^{min(m,n)} (-1)^{n_1}
    D_{m,n}^{n_1, n_2} \left({2\tilde{\vare}\Lambda\over |\tilde{v}^\perp_f|} \right) \,,
    \label{eq:Dmn}
    \ee
with $n_1(j)=j+(|m-n|-(m-n))/2$, $n_2(j)=j+(|m-n|+(m-n))/2$ in each term and
    \be
    D^{n_1,n_2}_{m,n}(z) =
    \left({-i\over\sqrt{2}}\right)^{n_1+n_2}
    {\sqrt{n!} \sqrt{m!} \over (n-n_1)! n_1! n_2!} W^{(n_1+n_2)}(z) \,,
    \ee
where $W^{(n)}(z)$ is the $n$-th derivative of the function $W(z)
= \exp(-z^2) \mbox{erfc}(-iz)$.
These functions have the following symmetries:
$W^{(n)}(z)^* = (-1)^n W^{(n)}(-z^*)$,
$\cD_{m,n} = (-1)^{m-n} \cD_{n,m}$ and
$D^{n_1,n_2}_{m,n}(z)^* = D^{n_1,n_2}_{m,n}(-z^*)$.

The diagonal part, $g$, is determined from the average $\ul{f} f$
and the normalization condition. The details, once
again, are relegated to Appendix \ref{app:F}, with the result
    \begin{subequations}
    \bea
    g &=& -i \pi / \sqrt{1+P} \,,
    \\
    P &=& -i\sqrt{\pi} {2\over w^2} \sum_n \sum_m \widetilde{\ul\Delta}_n \widetilde{\Delta}_m
    \sum'_{k,l\ge0} \frac{(\tilde{v}_+)^l(-\tilde{v}_-)^k}{l! \, k!}
    \nonumber \\
    &&\hspace{-1cm} \times \bra{n} a^{\dag k} a^l \ket{m} \left( {-i\over \sqrt{2}} \right)^{k+l}
    W^{(k+l+1)}\left( {\sqrt{2}\tilde{\vare}\over w}\right) \,,
    \eea
    \label{Eq:g-res}
    \end{subequations}
where $w=|\tilde{v}^\perp_f|/ \sqrt{2}\Lambda$, and the prime over
the $k,l$-sum denotes the restriction that the matrix element
$\bra{n} a^{\dag k} a^l \ket{m} = \sqrt{n!m!/(n-k)!(m-l)!}$ is
non-zero only for $k\le n$, $l\le m$ and $k-l=n-m$.

If we truncate the expansion of the order parameter in the vortex
state at the lowest Landau level function, $n=0$, we find from
Eqs.(\ref{Eq:g-res})
    \be
    g = \frac{-i \pi}{
    \sqrt{1-i\sqrt{\pi}\left(\frac{2\Lambda}{|\tilde{v}_f^\perp|}\right)^2
    \, W^\prime( \frac{2\tilde{\vare}\Lambda}{|\tilde{v}_f^\perp|} ) \,
    \widetilde\Delta_0 \ul{\widetilde\Delta}_0 } } \,,
    \label{eq:g0}
    \ee
which agrees with previously obtained
expressions\cite{WPesch:1975,PKlimesch:1978,AHoughton:1998,IVekhter:1999,HKusunose:2004}.
In the zero field limit, $\Lambda \propto 1/\sqrt{H} \to \infty$,
we use the asymptotic behavior at large values of the argument,
$W(z) \approx i/\sqrt\pi z$, $W^\prime(z)\approx -i/(\sqrt\pi
z^2)$, to verify that this Green's function tends to the BCS limit
\cite{AHoughton:1998,IVekhter:1999}, and therefore all the
conventional results for the density of states in nodal
superconductors immediately follow.

Eqs.(\ref{Eq:f-res}) and (\ref{Eq:g-res}) give the solution of the
quasiclassical equations in the BPT approximation for a given
vortex lattice and impurity self-energies,
i.e. provided the coefficients
$\Delta_n, \ul\Delta_m$ and $\Sigma, \Delta_{imp,n}, \ul\Delta_{imp,m}$
are known. The self-consistency equations for these coefficients,
    \bea
    %{1\over V_s} \Delta_n &=& \int {d\vare \over 4\pi i}
    &&\Delta_n \ln {T\over T_{c0}} = \int {d\vare \over 4\pi i}
    \int d\hvp_\sm{FS} \; n_f(\hvp) \, \cY(\hvp) \, \times
    \\
    &&\times \left( f^R_n(\hvp; \vare) - \ul{f}^R_n(\hvp; \vare)^*
    -2\pi i {\Delta_n \cY(\hvp)\over \vare} \right)
    \tanh {\vare\over 2 T} \,, \nonumber
    \eea
and the equations for the impurity retarded and advanced
self-energy, Eq.~(\ref{Eq:sigma-imp}), written explicitly through
solution of Eq.~(\ref{Eq:t-matrix}) for the $t$-matrix,
\begin{widetext}
    \be
    \widehat{t} = \left( \begin{array}{cc}
    t_+ + t_- & t_\Delta \, i\sigma_2 \\
    i\sigma_2 \, \ul{t}_\Delta & t_+-t_-
    \end{array} \right)
    =
    {1\over n_{imp}} \frac{\Gamma \sin^2\delta_0}{1-{\sin^2\delta_0\over \pi^2}
    (\langle g\rangle^2-\overline{\langle f\rangle \langle \ul{f} \rangle} + \pi^2 )}
    \left( \begin{array}{cc}
    \cot\delta_0 + \langle g\rangle/\pi & (\langle f\rangle /\pi)  \, i\sigma_2 \\
    i\sigma_2 \, (\langle \ul{f} \rangle/\pi) & \cot\delta_0 - \langle g\rangle/\pi
    \end{array} \right) \,,
    \label{eq:explicit_imp}
    \ee
\end{widetext}
complete the closed form solution. Here $T_{c0}$ is the critical
temperature for the clean system, $\Gamma=0$, which we used to
eliminate the interaction strength, $V_s$, and the high energy
cutoff. The elimination can also be done in favor of the impurity
suppressed $T_c$, see e.g. Ref.\onlinecite{xu95}.

\section{Heat capacity}
\label{sec:Thermo}

\subsection{Density of states and the specific heat}

Once we self-consistently determined the Green's function, we can
calculate the quasiparticle spectrum. We use the standard
definition for the angle-resolved density of states at the Fermi
surface,
    \be
    \frac{N(\vare, \hvp)}{N_f(\hvp)} = -{1\over \pi} \Im g^R(\hvp, \vare)
    \,,
    \ee
where $N_f$ is the normal state DOS.

The heat capacity is the derivative of the entropy, $C = T \,
\partial S/\partial T$, where
    \bea
    S &=& - 2 \sum_\vk [(1-f(E_\vk))\ln(1-f(E_\vk)) + f(E_\vk) \ln f(E_\vk)]
    \nonumber \\
    &=& - 2 \int\limits_{-\infty}^{+\infty}d\vare \, N(\vare) \,
    [(1-f(\vare))\ln(1-f(\vare)) + f(\vare) \ln f(\vare)] \,,
    \nonumber
    \eea
$f(\vare) = 1/(e^{\vare/T}+1)$ is the Fermi function, and
$N(\vare) = \int d\hvp \, N(\vare, \hvp)$ is the net DOS at energy
$\vare$. In practice, numerical differentiation of the entropy is
computationally either noisy or very time consuming due to high
accuracy required in finding $S$, and therefore not very
convenient. At low temperatures the order parameter and the
density of states are weakly temperature dependent, and therefore
the specific heat can be obtained by differentiating only the
Fermi functions. This leads to the well-known expression
    \begin{equation}
    C(T,\bm H) = \frac{1}{2}  \int\limits^{+\infty}_{-\infty}
    \; d\vare \; \frac{\vare^2 \; N(T, \bm H; \vare)}{T^2
    \cosh^2(\vare/2T)} \,,
    \label{eq:C}
    \end{equation}
that lends itself more efficiently to numerical work. Note that
the $x^2/\cosh^2(x/2)$ function has a single sharp peak at
$x\sim2.5$, so the DOS at $\vare \sim 2.5\div 3 \, T$ contributes
the most to the $C/T$. The difference between the specific heat
defined from the density of state and the exact result is, of
course, dramatic near the phase transition from the normal metal
to a superconductor, where the peak in the specific heat is
entirely due to entropy change not accounted for in
Eq.~(\ref{eq:C}). At the same time, the regime where the
anisotropy of $C(T,\bm H)$ is measured is far from $T_c$, and
there we find that the results are very weakly dependent on the
method of calculation. We therefore use the approximate expression
above except where noted, and give a more detailed account of the
difference between the two approaches for the specific Fermi
surface shape in Sec.~\ref{sec:QCYL}.

%~~~~~~~~~~~~~~~~~~~~~~~~~~~~~~~~~~~~~~~~~~~~~~~~~~~~~~~~~~~~~~~~~~~~~~~~~
\subsection{\label{sec:CYL} Cylindrical Fermi surface}
%~~~~~~~~~~~~~~~~~~~~~~~~~~~~~~~~~~~~~~~~~~~~~~~~~~~~~~~~~~~~~~~~~~~~~~~~~

We are now prepared to consider the behavior of the specific heat
in the vortex state of a superconductor. As mentioned above, our
goal is to analyze the variations of the specific heat when the
applied field is rotated with respect to the nodal directions. We
consider first the simplest model of a cylindrical Fermi surface
with vertical lines of nodes, and  the field applied in the basal
plane, at varying angle to the crystal axes.

This is a simplified version of a model for layered compounds,
such as CeCoIn$_5$, considered below in Sec.~\ref{sec:QCYL}. There
we compute the specific heat for a quasi-cylindrical Fermi
surface, open and modulated along the $z_0$-axis. The main
advantage of considering an uncorrugated cylinder first is that it
provides a good basis for semi-analytical understanding of the
main features of the thermodynamic properties. Moreover, this
model gives results that are in semi-quantitative agreement with
those for the more realistic model of Sec.~\ref{sec:QCYL}.

The disadvantage of the model is that it is not self-consistent.
If the Fermi surface is cylindrical, there is no component of the
quasiparticle velocity along the $z_0$ direction (the axis of the
cylinder). The field applied in the plane does not result in the
Abrikosov vortex state, as the supercurrents cannot flow between
the layers. Consequently, it is impossible to set up and solve the
self-consistency equations for the order parameter as a function
of the applied field. Nonetheless we assume the existence of the
vortex lattice where the order parameter has a single $n=0$ Landau
level component, with the amplitude $\Delta(T,H) = \Delta(T,0)
\sqrt{1-H/H_{c2}(T)}$, analogous to
Ref.\onlinecite{IVekhter:1999,IVekhter:1999R}.  With this
assumption, we solve self-consistently for the
temperature-dependent $\Delta(T,0)$, and for the impurity
self-energies. We consider the unitarity limit of impurity
scattering (phase shift $\delta_0=\pi/2$). In the next section we
compare this model with a more realistic fully self-consistent
approach, and show that the major features of the two are very
similar.

While in the cylindrical approximation the results depend solely
on the ratio $H/H_{c2}$, for comparison with the results of the
self-consistent calculation we recast them in similar form. We
measure the field in the units of $B_0 = \Phi_0 / 2\pi \xi_0^2$
where $\Phi_0=hc/2|e|$ is the flux quantum and $\xi_0 = \hbar
v_f/2\pi T_c$ is the temperature independent coherence length in
the $ab$-plane. At zero temperature the upper critical field along
the $c$-axis is computed self-consistently, $H_{c2,c} \approx 0.55
B_0$. We set the in-plane $H_{c2} = 1.1 B_0$ to approximate the
factor of 2 anisotropy found in CeCoIn$_5$, and choose the normal
state scattering rate $\Gamma/2\pi T_c=0.007$ (suppression of the
critical temperature $(T_{c0}-T_c)/T_{c0} \approx 5\%$). We
checked that the resulting map of the anisotropy in the specific
heat in the $T$-$H$ plane does not strongly depend on this
particular choice. Of course, large impurity scattering smears the
angular variations.

\begin{figure}[t]
\centerline{\includegraphics[height=5.5cm]{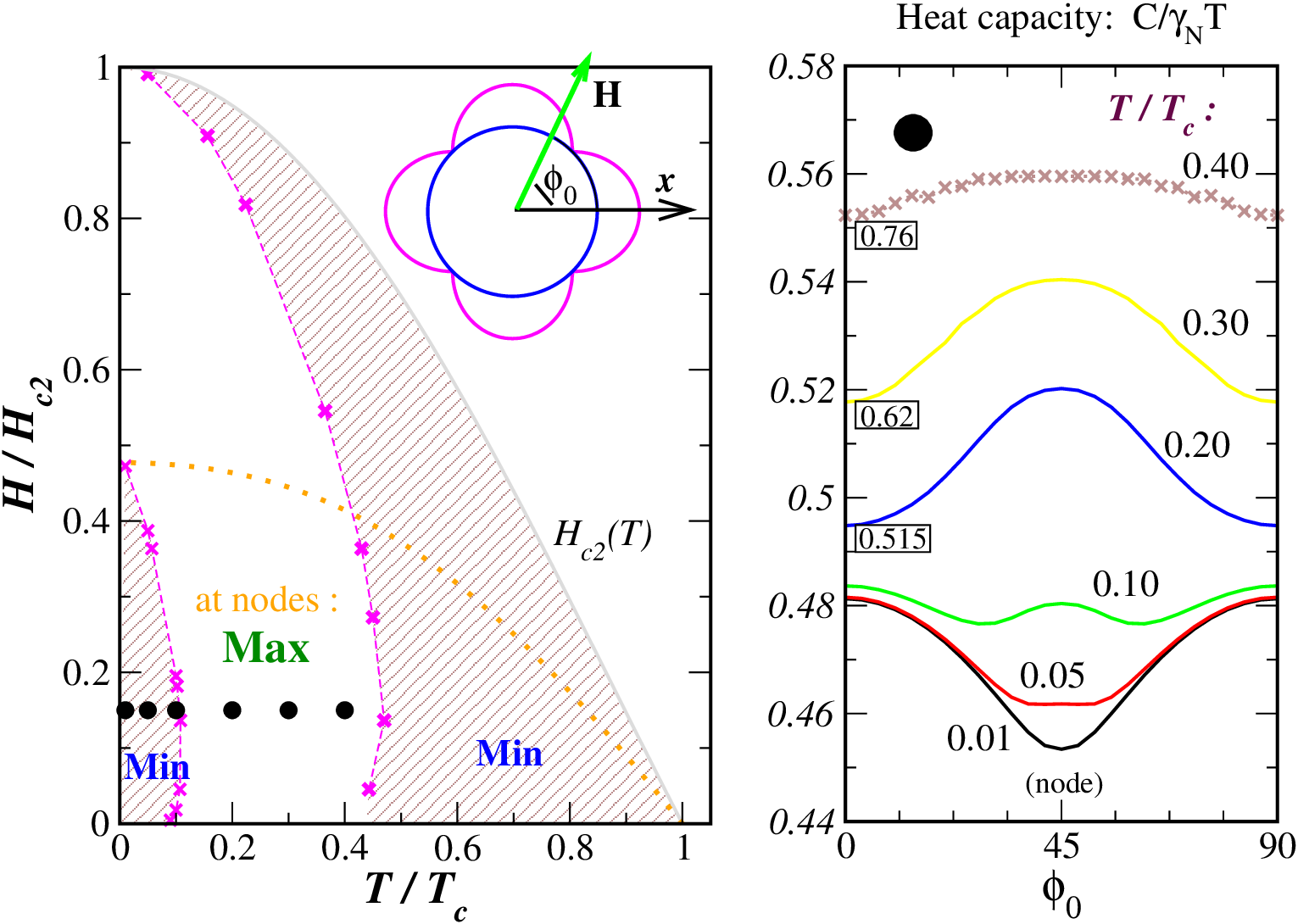}}
\caption{\label{fig:PDcylC} (Color online)
Left panel:
The phase diagram for the anisotropy of the heat capacity for cylindrical
Fermi surface. At low $T$ and $H$ (shaded area) the minimum in the
heat capacity occurs when the field point in a nodal direction,
$\phi_0=45$. As $T$ increases the minimum first evolves into a
maximum, and then switches back to a  minimum. The inversion of
zero energy DOS is indicated by the dotted line.
Right panel:
evolution of the heat capacity anisotropy with temperature for
$H/H_{c2} = 0.136$ (circles in the left panel).
Some curves are shifted vertically for clarity,
their original values at $\phi_0=0$ are shown in boxes.
$\gamma_N$ is the Sommerfeld coefficient in the normal state.
}
\end{figure}

For a single Landau level component the solutions for the Green's
function have a particularly simple form of Eq.~(\ref{eq:g0}). For
a $d_{x^2-y^2}$ superconductor the gap function is
$\Delta(\phi)=\Delta\cos 2\phi$. If the magnetic field is applied
at an angle $\phi_0$ to the $x$-axis (inset in
Fig.~\ref{fig:PDcylC}), the component of the Fermi velocity normal
to the field is
\begin{equation}
  v_f^\perp(\phi)=v_f\sin(\phi-\phi_0).
\end{equation}
Therefore the Green's function of Eq.(\ref{eq:g0}) takes the form
\begin{widetext}
\begin{equation}
  g (\vare,\phi) =\frac{-i \pi}{
    \sqrt{1-i\sqrt{\pi}\left(\frac{2\Lambda\Delta}{v_f|\sin(\phi-\phi_0)|}\right)^2
    \, W^\prime( \frac{2\tilde{\vare}\Lambda}{v_f|\sin(\phi-\phi_0)|} ) \cos^2 2\phi } }
    \,
    \label{eq:g-one}
\end{equation}
\end{widetext}

Let us focus first on the residual density of states, $\vare\rightarrow 0^+$
in the clean limit, $\Gamma=0$, to compare with the semiclassical Doppler approximation.
In this case $W^\prime(0)=2i/\sqrt\pi$ and the density of states reduces to
    \be
    N(0) = \int_0^{2\pi} {d\phi\over 2\pi}
    \frac{1}{\sqrt{1+{1\over4z^2}{\cos^2 2\phi\over\sin^2(\phi-\phi_0)}}}
    \label{eq:ZeDosAnal}
    \ee
where $z=v_f/4\sqrt{2}\Lambda\Delta \sim \sqrt{H/H_{c2}}$. The DOS
can be obtained analytically for the nodal and antinodal
alignments of the field.

{\em Node, $\phi_0=\pi/4$.} Then Eq.(\ref{eq:ZeDosAnal}) reduces
to
    \bea
    N_\sm{node}(0) = {2\over \pi} {z\over\sqrt{1+z^2}} K\left( {1\over\sqrt{1+z^2}} \right) \,,
    \label{eq:N0node}
    \eea
where $K$ is the complete elliptic integral of the first kind. We
use the convention of Ref.~\onlinecite{GR} for the argument of all
elliptic functions. In the weak field limit, $z\ll 1$,
    \be
    N_\sm{node}(0) \simeq {2z\over\pi} \ln {4\over z} \,.
    \ee

{\em Antinode, $\phi_0=0$.} The corresponding DOS is evaluated to
be
    \begin{subequations}
    \label{eq:N0anode}
    \be
    N_\sm{antinode}(0) = {z\over (z^2+1/4)^{1/4}} {2\over\pi}
    \left[ K(r) - {1\over2} F(\alpha, r) \right] \,,
    \ee
    where $F(\alpha,r)$ is the incomplete elliptic integral of the first kind, and
    \be
    \alpha = \arccos \frac{1-\sqrt{z^2+1/4}}{1+\sqrt{z^2+1/4}}
    \; , \;
    r= {1\over \sqrt{2}} \sqrt{1+{1 + z^2 \over \sqrt{1+4z^2}}} \,.
    \ee
    \end{subequations}
At low fields, $z\ll 1$, the antinodal DOS
    \be N_\sm{antinode}(0)
    = {2\sqrt{2} z\over\pi} \ln {4\sqrt{2}\over z} \,.
    \ee
Apart from the logarithmic correction (which is rapidly washed
away by finite impurity scattering), the antinodal DOS exceeds the
nodal value by a factor $\sqrt{2}$, in complete agreement with the
Doppler approach \cite{IVekhter:1999R}. As the field increases
however, Eqs. (\ref{eq:N0node}) and (\ref{eq:N0anode}) predict a
crossing point $z^* \sim 0.63$ above which the residual nodal DOS
becomes greater than $N_\sm{antinode}(0)$; this result was
obtained numerically in
Refs.~\onlinecite{MUdagawa:2004,PMiranovic:2003}. With our choice
of $\Delta(H) = \Delta \sqrt{1-H/H_{c2}}$, the zero-temperature
crossover point lies at $H^*/H_{c2} \sim 0.6$.

Similar analytic expressions cannot be written for finite energies
and we evaluate the DOS and the specific heat numerically,
including the impurity effects. Results for the anisotropy of the
heat capacity are shown in Fig.~\ref{fig:PDcylC}. We present them
in a form of a phase diagram (left panel) that shows the regions
with the opposite anisotropy. Shaded (white) areas correspond to
the minimum (maximum) of $C$ when $\vH$ is along a node.
 Of course the
node-antinode anisotropy disappears as the field $H\to 0$. Since
we are primarily interested in comparison of our results with the
experimental data, we focus on the regime of moderate fields, and
show the evolution of specific heat for different directions of
the field, $\phi_0$, with the temperature in the right panel of
Fig.~\ref{fig:PDcylC}. Notice that at $\phi_0=45^\circ$, when the
field is along a nodal direction, the minimum in $C(\phi_0)$
evolves into the maximum as $T$ increases.

%%%%%%%%%%%%%%%%%%%%%%%%%%%%%%%figure
%%%%%%%%%%%%%%%%%%%%%%%%%%%%%%%figure

Inversion of the anisotropy in the $T$-$H$ phase diagram is at
odds with the semiclassical result that always predicts a minimum
in the specific heat for the field parallel to the nodal
direction. In the shaded area adjacent to the $H_{c2}(T)$ line in
Fig.~\ref{fig:PDcylC}, with minima for $\vH || \, node$, the
specific heat is already sensitive to the density of states near
the BCS singularity in the DOS at $\vare \sim \Delta_0$, and
therefore direct comparison with the semiclassical analysis is not
possible. Moreover, we show in the following section that the
self-consistent models require nodal-antinodal anisotropy of the
upper critical field, and the results for this part of the phase
diagram are modified.

\begin{figure}[t]
\centerline{\includegraphics[height=6.0cm]{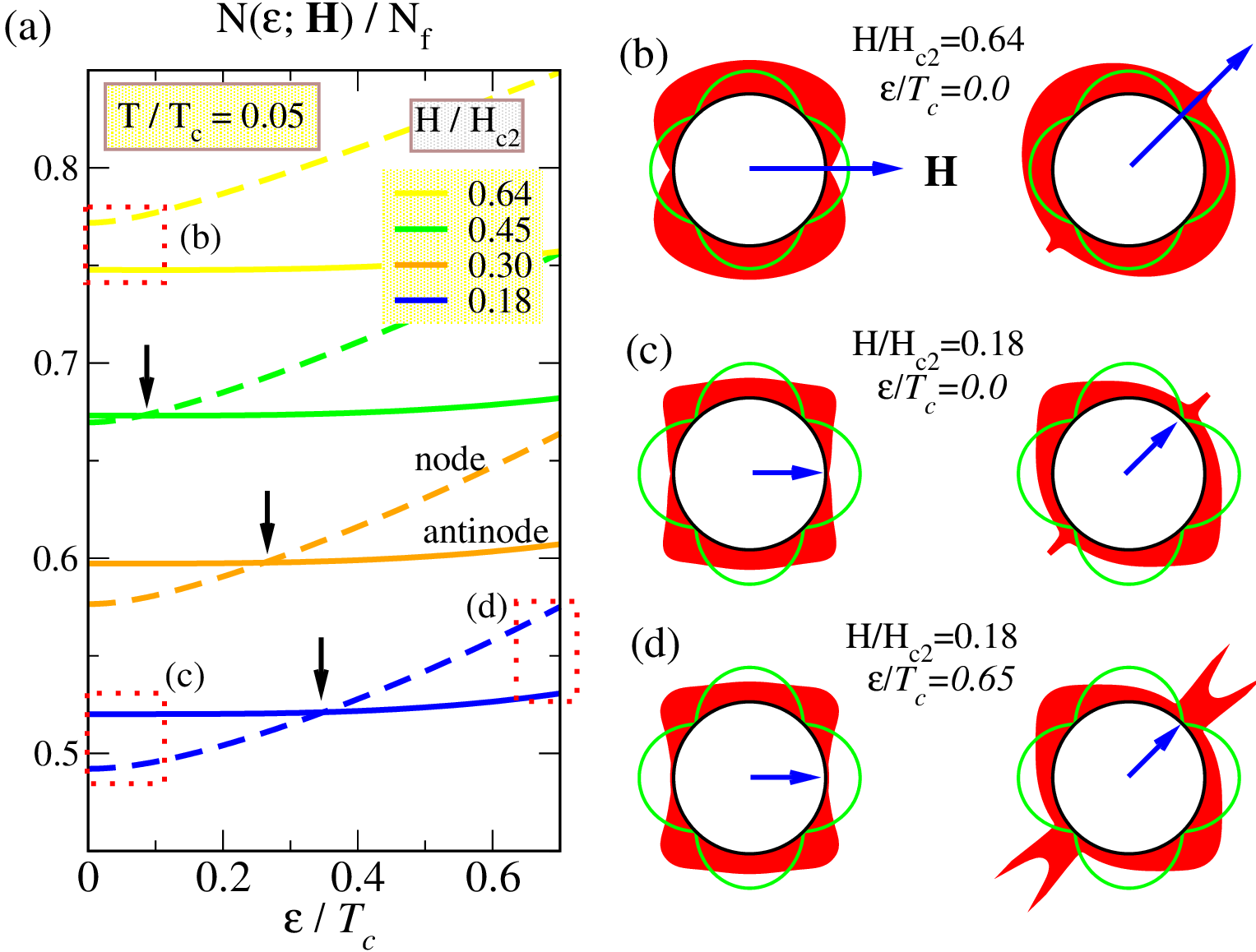}}
\caption{\label{fig:doscyl} (Color online) Left: the low-energy
part of the DOS for cylindrical FS. The nodal and antinodal DOS
cross at finite energies (arrows). Right: the angle-resolved DOS
(red shaded) for the two field orientations in the regions
indicated by the dotted boxes in the left panel. The angle
integrated DOS is given by the area of the shaded regions. See
text for details. }
\end{figure}

On the other hand, the anisotropy inversion between the low-$T$,
low-$H$ region, and the intermediate temperatures and fields,
occurs still in the regime where the semiclassical logic may have
been expected to work. The dotted line in the left panel of
Fig.~\ref{fig:PDcylC} separates the two regions of the residual
zero-energy DOS: below that line $N_\sm{node}(0) <
N_\sm{antinode}(0)$, while above the line $N_\sm{node}(0) >
N_\sm{antinode}(0)$. The inversion of the anisotropy in the
specific heat is clearly not just a consequence of the behavior of
the zero-energy DOS found above. Recalling that $C/T$ is
predominantly sensitive to the density of states at energies of
the order of a few times $T$ (see Eq.(\ref{eq:C})), we conclude
that the origin of the anisotropy inversion is in the behavior of
the finite energy DOS. We plot the low-energy $N(\vare,{\bf H})$
at several values of the magnetic field in the left panel of
Fig.~\ref{fig:doscyl}. At low fields, the  DOS anisotropy at small
$\vare$ agrees with the semiclassical prediction, but the density
of states for the field along a node (dashed lines) and along an
antinode (solid lines) become equal at a finite energy indicated
by arrows. Above this energy the DOS anisotropy is reversed, and
is manifested in the reversal of the specific heat anisotropy as
$T$ increases. The crossing point moves to lower energies with
increasing field, and is driven to zero when the residual,
$\vare=0$, DOS for the two directions become equal. In our
numerical work with finite impurity scattering rate this occurs at
$H^* \sim 0.5 H_{c2}$, and we checked that $H^*\to 0.6 H_{c2}$ as
the system becomes more pure, in agreement with the analytical
results above.

%%%%%%%%%%%%%%%%%%%%%%%%%%%%%%%figure
%%%%%%%%%%%%%%%%%%%%%%%%%%%%%%%figure

As suggested by us in the short Letter communicating our main
results, the inversion stems from the interplay between the energy
shift and scattering due to magnetic field \cite{AVorontsov:2006}.
Magnetic field not only creates new quasiparticle states on the
Fermi surface, but also scatters the quasiparticles and,
consequently, re-distributes their spectral density. This
scattering is present in the microscopic method, but not in the
Doppler shift treatment. To understand this effect and to make
connection with the semiclassical approach we analyze the
angle-resolved DOS obtained from the Green's function,
Eq.~(\ref{eq:g0}). It is instructive to re-write the Green's
function in the BCS-like form which makes explicit the distinction
between the energy shift and scattering rate. We define the
``magnetic self-energy'' $\Sigma=\Sigma'-i\Sigma''$ from $(\vare -
\Sigma)^{-2} \equiv i\sqrt\pi (2\Lambda/ |\tilde{v}_f^\perp|)^2 \,
W^\prime(2\tilde\vare\Lambda/|\tilde{v}_f^\perp|)$ so that the
Green's function reads
    \be
    g^R = -i \pi
    \left(1- \frac{|\Delta_0|^2 \cY^2(\hat\vp) }
    {(\vare - \Sigma'(\vare,\vH,\hvp)+i\Sigma''(\vare,\vH,\hvp))^2} \right)^{-1/2} \,.
    \label{eq:magsigmas}
    \ee
The density of states for a given direction at the Fermi surface
can be found from the comparison of $\vare-\Sigma(\hvp)$ with
$\Delta_0 \cY(\hat\vp)=\Delta_{max} \cos 2\phi$. Since $W'(x)$ is
a complex-valued function, both $\Sigma'$ and $\Sigma''$ are
generally non-zero: the former shifts the quasiparticle energy,
while the latter accounts for the direction-dependent scattering.
For now we neglect the impurity broadening: for quasiparticles
moving not too close to the field direction the field-induced
scattering is normally greater than the scattering by impurities.
Non-zero $\Sigma''$ is the key signature of our microscopic
solution. Both real and imaginary components of the self-energy
depend on the quasiparticle energy $\vare$, the strength and
direction of the field $\vH$ and on the momentum of the
quasiparticle with respect to both nodal direction and the field.
Using the expansion around $W'(0) = 2i/\sqrt\pi$ at small values
of the argument, and taking $W'(z\gg 1) \approx -i/\sqrt\pi z^2$
for large arguments, we find two limiting cases,
    \bea
    &&\vare - \Sigma
    \approx i{|\tilde{v}_f^\perp|\over 2\sqrt{2}\Lambda} +
    O\left(\frac{\Lambda^2\vare^2}{|\tilde{v}_f^\perp|^2}\right),\;
    \mbox{if }  \vare \ll \frac{|\tilde{v}_f^\perp|}{2 \Lambda } \;,
    \label{eq:lowExp}
\\
    &&\vare - \Sigma  \approx \vare
    +O\left(\frac{|\tilde{v}_f^\perp|^2}{\Lambda^2\vare^2
    }\right),\qquad\; \; \;
    \mbox{if }\vare \gg \frac{|\tilde{v}_f^\perp|}{2 \Lambda } \,.
    \eea
Note that $|\tilde{v}_f^\perp (\widehat{\bf p})| / 2\Lambda\propto
\sqrt{H}$ is the characteristic magnetic energy scale for
quasiparticles at position $\widehat{\bf p}$ on the Fermi surface.
In the first limit, valid at low energies (or moderately strong
fields) for quasiparticle momenta away from the field direction,
the imaginary part of the self-energy is dominant. In the opposite
limit the effect of the field is small. Between these two limits,
\ie at finite energies, moderate fields and arbitrary $\hvp$ , the
real (energy shift) and the imaginary (scattering) parts of the
self-energy can be comparable.

We can now analyse the angle-dependent contribution to the density
of states from different regions at the Fermi surface at a given
field, which is shown in the right panel of Fig.~\ref{fig:doscyl}.
Consider first very low energy $\vare \to 0$, panels Fig.~\ref{fig:doscyl} b) and c),
so that we are in the regime described by Eq.(\ref{eq:lowExp}). At
low fields, panel c), the characteristic energy,
$|\tilde{v}_f^\perp (\widehat{\bf p})| / 2\Lambda$ is smaller than
the maximal gap, $\Delta_{max}$, and therefore most of the field-induced
quasiparticle states appear near the nodes
for which $|\tilde{v}_f^\perp (\widehat{\bf p})|$ is moderately large.
Consequently, as in the semiclassical result, field applied along
a nodal direction does not create quasiparticles near that node,
while the field applied along the gap maximum generates new states
at all nodes. The small contribution seen in the
right frame of panel c) at the nodes aligned with the field is due
to impurity scattering.
Thus, while the  scattering on the vortices, \ie the
imaginary part of Eq.(\ref{eq:lowExp}), does produce a
non-vanishing contribution to the DOS over most of the Fermi
surface, at very low energy and low field
the spectral weight of the field-induced
states is mainly concentrated near the nodal points.

This changes as the field is increased, see panel b). At high
field the Doppler shift and pairbreaking due to scattering are
strong, and sufficient to contribute to the single particle DOS
over almost the entire Fermi surface where $|\tilde{v}_f^\perp
(\widehat{\bf p})| / 2\Lambda\sim\Delta_{max}(H)$ (as a reminder,
in our notations the maximal gap, $\Delta_{max}=\Delta_0\sqrt{2}$,
since we chose $\cY(\phi)=\sqrt{2}\cos2\phi$ to be normalized).
The obvious exceptions are the momenta close to the direction of
the field, when $|\tilde{v}_f^\perp (\widehat{\bf p})|\ll v_f$.
For the field aligned with the node this restriction is not
severe: near the node ${v}_f^\perp\simeq v_f\delta\phi$ and
$\Delta(\hvp)\simeq 2\Delta_{max}\delta\phi$, where $\delta\phi$
is the deviation from the nodal (and field) direction. Hence if
$v_f/(2\Lambda)\sim\Delta_{max}$, almost the entire Fermi surface
contributes to the DOS when the field is aligned with a node. In
contrast, for the field along the gap maximum, in a range of
angles close to the field direction the gap is large and the
magnetic self-energy is small, and hence no spectral weight is
generated. As a result, the density of states is higher for the
nodal orientation. This is the origin of zero-energy DOS inversion
as found numerically in Ref.~\onlinecite{MUdagawa:2004} and as
derived above.

We finally consider the DOS at finite energy
$\vare\ll\Delta_{max}$, panel (d). In the absence of the field the
most significant contribution to $N(\vare)$ comes from the BCS
peaks at $\vare=\Delta_0|\cY(\hat\vp)|$, located at momenta
$\hvp_\vare$ at angles $\phi_n\pm\delta\phi_\vare$, where
$\phi_n=\pi/4+\pi n/2$ are the nodal angles, and
$\delta\phi_\vare\approx\vare/(2\Delta_{max})=\vare/(2\sqrt{2}\Delta_0)$.
Scattering on impurities or vortices broadens these peaks, and
re-distributes the spectral density to different energies (as in
all unconventional superconductors, scattering reduces the weight
of the singularity and piles up spectral weight at low energies).
However, the vortex scattering is anisotropic as it depends on
${v}_f^\perp$, see Eq.(\ref{eq:lowExp}), the component of the
velocity normal to the field. Therefore if a field is applied
along a nodal direction, at that node ${v}_f^\perp\simeq
v_f\delta\phi_\vare\ll v_f$, and the peaks in the angle resolved
DOS remain largely intact (d,right). On the other hand, if the
field is applied along a gap maximum, BCS peaks near all four
nodes are broadened by scattering, and their contribution to the
net DOS is reduced (d,left). So, even when the field is moderately
low but the quasiparticle energy exceeds some value $\vare^\star$,
which can only be determined numerically, the gain from sharp
(unbroadened by scattering) coherence peaks exceeds the
field-induced contribution from the near-nodal regions. Then the
DOS is higher for the field along a node rather than the gap
maximum. Recalling that the specific heat at temperature $T$ is
largely controlled by the density of states at the energy of about
$2.5 T$, we expect that the anisotropy of the specific heat is
also inverted at $T/T_c \sim \vare^\star/2.5 T_c$. It is this
change in the {\em finite-energy} density of states, rather than
the zero energy DOS, that determines the inversion line in the
phase diagram, see Fig.~\ref{fig:PDcylC}.

\subsection{\label{sec:QCYL} Quasi-two-dimensional Fermi surface}

We mentioned above that a major motivation of our work is to
address the apparent discrepancy between the thermal conductivity
and specific heat measurements in CeCoIn$_5$. While this material
does possess a quasi-two dimensional sheet of the Fermi surface,
the normal state resistivity anisotropy is very moderate,
indicating a significant $c$-axis electronic dispersion.
Consequently, while the results of the previous section are very
suggestive of the anisotropy reversal, we need to verify that
similar physics persists in a more realistic open
quasi-cylindrical Fermi surface, described by
$$p_f^2 = p_x^2 + p_y^2 - (r^2 \, p_f^2) \cos (2 s\, p_z/r^2 p_f) \,.$$
We parameterize this FS by the azimuthal angle in the $ab$-plane,
$\phi$,  and momentum along the $c$-axis, $p_z$, so that the Fermi
velocity at a point $(\phi,p_z)$ is
$$\vv_f(p_{z}, \phi) = \left( \begin{array}{c}
p_f/m \; \sqrt{1+r^2 \cos(2 s p_{z}/r^2 p_f)} \cos\phi  \\
p_f/m \; \sqrt{1+r^2 \cos(2 s p_{z}/r^2 p_f)} \sin\phi  \\
p_f s/m \; \sin(2 s p_{z}/r^2 p_f)
\end{array} \right) \,.$$
With this parametrization, the anisotropy factor of the normal
state DOS is $n_f(\hvp) = 1$. Parameter $r$ determines the
corrugation amplitude along the $z$-axis, and we find that the
results do not depend on its value;  below we set $r=0.5$. The
second parameter, $s$, is physically important since it fixes the
anisotropies of the normal state transport and the critical field:
the characteristic velocities in the $ab$-plane and along the
$c$-axis are $v_{0\perp} = p_f/m \equiv v_f$ and $v_{0||} = p_f s
/m$. The normal state conductivity anisotropy is
$\sigma_{zz}/\sigma_{xx}=s^2$.

%%%%%%%%%%%%%%%%%%%%%%%%%%%%%%%figure
\begin{figure}[t]
\centerline{\includegraphics[height=5.5cm]{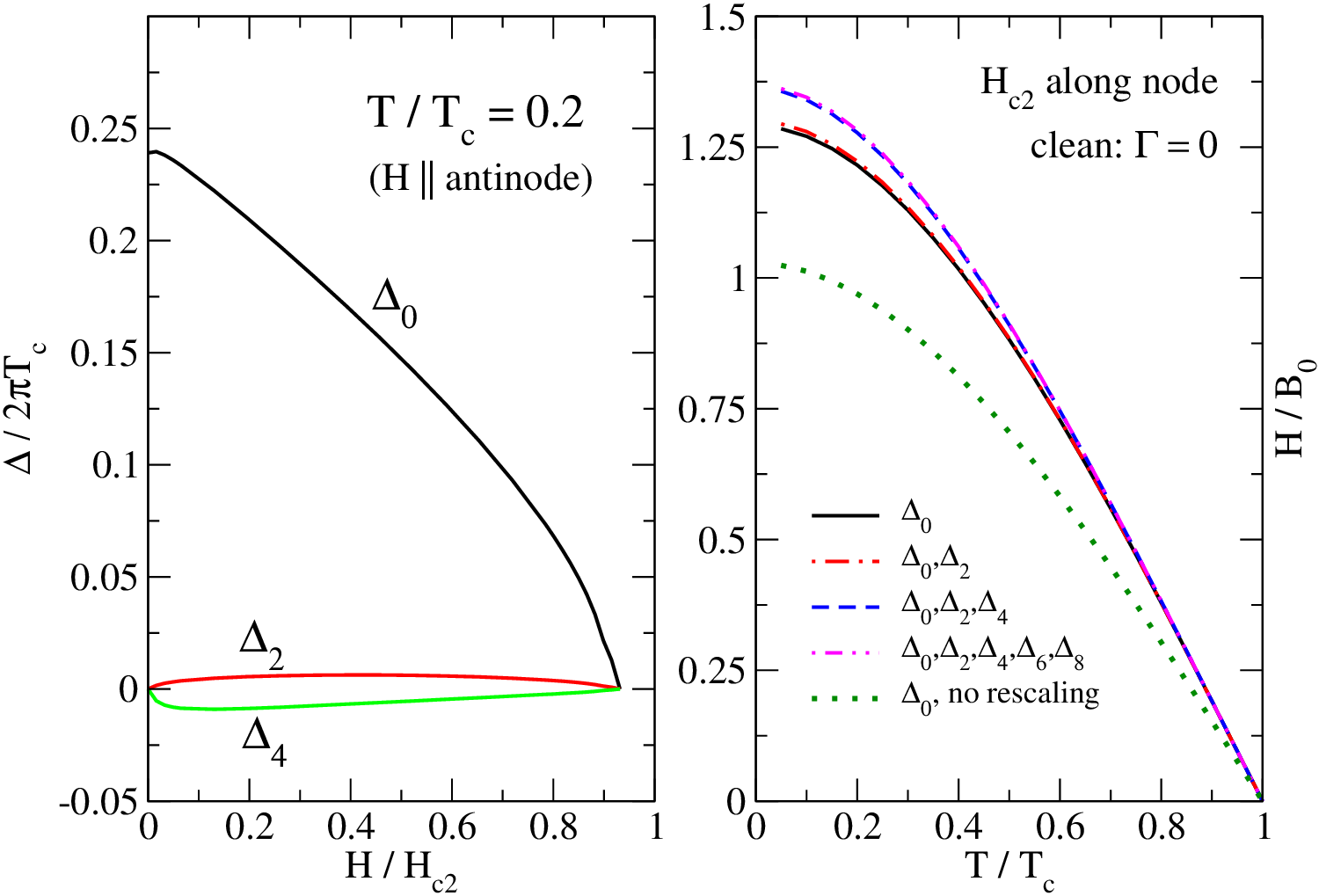}}
\caption{\label{fig:op} (Color online) Multiple Landau level
contributions to the order parameter. Left panel: LL components as
a function of the field in the antinodal direction for $T/T_c=0.2$
(left). Right panel: upper critical field along a node for
different number of Landau channels in $\Delta$ (clean limit);
$H_{c2}$ has converged for $N\ge3$. We also show for comparison
the $H_{c2}$ as calculated with one channel $\Delta_0$ without
coordinate rescaling (\ref{eq:xyscale}). }
\end{figure}
%%%%%%%%%%%%%%%%%%%%%%%%%%%%%%%figure

The main advantage of allowing for the $c$-axis dispersion is the
ability to solve the quasiclassical equations in the BPT
approximation self-consistently, with respect to both the order
parameter as a function of $T,H$, and the impurity self-energy. We
take moderate values of the anisotropy, $s=0.25$ and $s=0.5$, for
which the vortex structure is still three-dimensional. The latter
value yields $H_{c2}$ anisotropy close to that of CeCoI$_5$. The
calculations below are done with three Landau level channels for
the order parameter, ${\Delta_0, \Delta_2, \Delta_4}$. With the
rescaling of Appendix \ref{app:GL} this is sufficient for
convergence of the upper critical field. The values of the higher
components $\Delta_2, \Delta_4$ are less than $5\%$ of $\Delta_0$,
see Fig.~\ref{fig:op}, and addition of further components does not
change the results.

For this Fermi surface we solve the linearized self-consistency
equation and compute $H_{c2}$ in the basal plane. The anisotropy
between nodal and antinodal upper critical fields appears
naturally as a result of the $d$-wave symmetry, $H_{c2}^{node}\ne
H_{c2}^{antinode}$. The value of $H_{c2}$ is essentially
determined by balancing the kinetic energy of the supercurrents
vs. the condensation energy, and the former is different for
different orientations of the field.

\begin{figure}[t]
\centerline{\includegraphics[height=5.5cm]{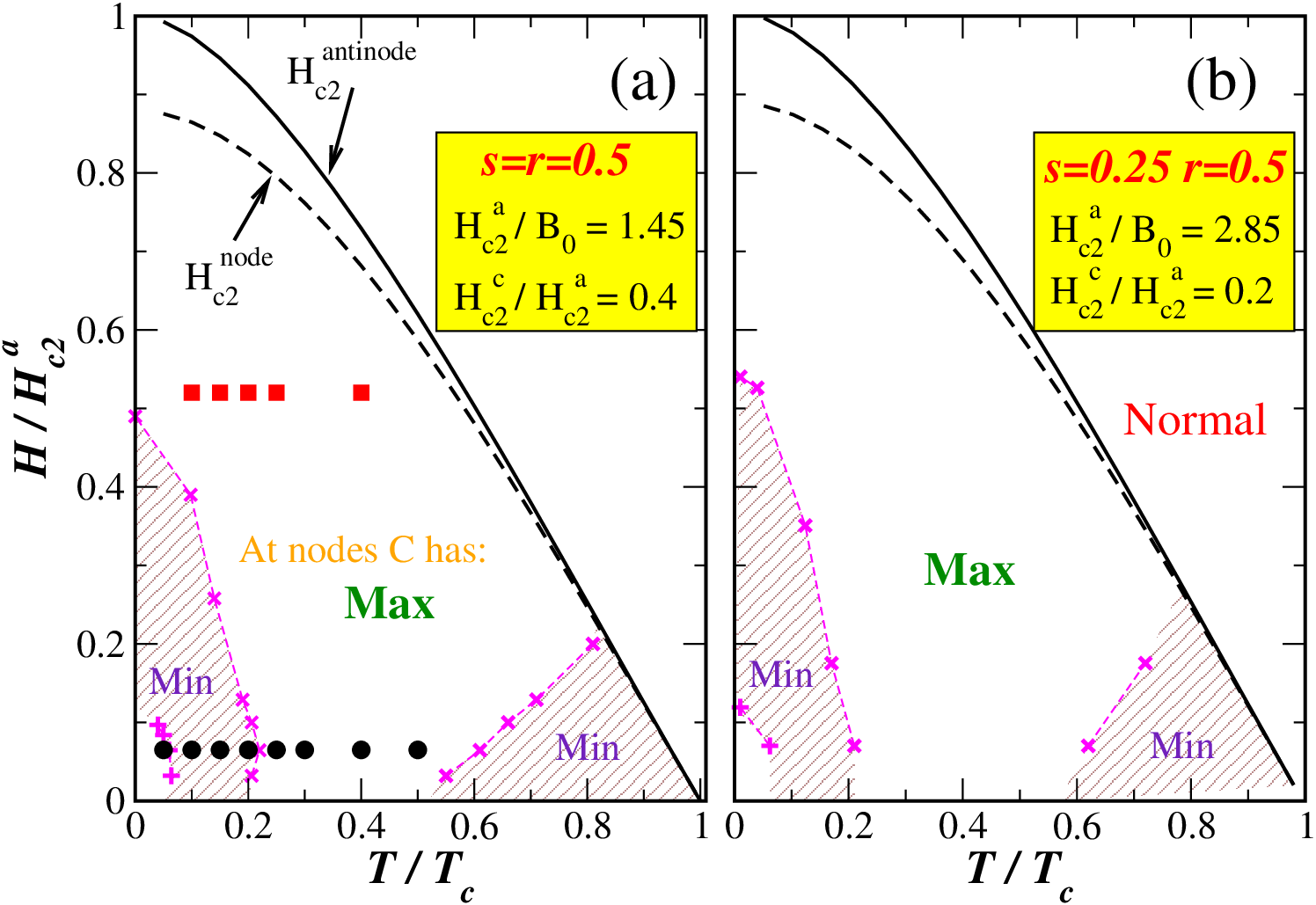}}
\caption{\label{fig:PDqcylC} (Color online) The phase diagram of
the heat capacity anisotropy $C(\phi_0)$ for the corrugated FS
with $s=r=0.5$ (left) and $s=r/2=0.25$ (right). }
\end{figure}

Let us now look at  the difference between the self-consistent and
non-self-consistent order parameter calculations. For this we
again present a phase diagram, Fig.~\ref{fig:PDqcylC}, analogous
to Fig.\ref{fig:PDcylC} for the cylindrical FS. Left panel shows
the results for the Fermi surface with $r=s=0.5$, and the impurity
strength $\Gamma/2\pi T_c = 0.007$ ($T_c/T_{c0} \sim 0.95$,
$\ell_{tr}/\xi_0 \simeq 70$). The values of the critical fields at
$T=0$ are $H_{c2}^{antinode} \approx 1.45 B_0$, $H_{c2}^{node}
\approx 1.27 B_0$ and $H_{c2}^{c} \approx 0.57 B_0$. This gives
the in-plane anisotropy
$(H_{c2}^{antinode}-H_{c2}^{node})/H_{c2}^{antinode} \sim 15\%$,
and the ratio between the $c$-axis and antinodal directions,
$H_{c2}^{c}/H_{c2}^{antinode}=0.4$.
%%%%%%%%%%%%%%%%%%%%%%%%%%%%%%figure
%%%%%%%%%%%%%%%%%%%%%%%%%%%%%%figure
To demonstrate the influence of the FS $c$-axis curvature on this
phase diagram, we present in Fig.~\ref{fig:PDqcylC}(b) a similar
diagram a Fermi surface with parameters, $r=0.5$ and $s=0.25$.
These parameters correspond to the reduction by factor of 2 of the
velocity along the $c$-axis and the critical fields
$H_{c2}^{antinode} \approx 2.85 B_0$, $H_{c2}^{node} \approx 2.55
B_0$ and $H_{c2}^{c} \approx 0.57 B_0$. The $H_{c2}$ anisotropies
are: 10\% in the basal plane between nodal and antinodal
directions, and $H_{c2}^{c}/H_{c2}^{antinode}=0.2$.
Fig.~\ref{fig:PDqcylC} shows that a factor of two difference in
the $c$-axis velocity affects only the absolute values of
$H_{c2}^{(anti)node}$, but otherwise the two diagrams for the
anisotropy in the $ab$-plane look almost identical.

The shaded ``semiclassical'' region at low temperatures and fields
in Fig.~\ref{fig:PDqcylC}, where minima of $C$ are for $\vH || \,
node$, expanded compared with that for cylindrical FS
(Fig.~\ref{fig:PDcylC}). We note that if we truncate the order
parameter expansion at the lowest Landau level, without full
convergence of $H_{c2}$, the ``nodal minimum'' region occupies
similar range for both corrugated and purely cylindrical FS.
Therefore this expanded range is the result of the
self-consistency and inclusion of higher harmonics. On the other
hand, the shaded `minimum-at-a-node' region near $H_{c2}$ shrunk
to low $H$ and high $T$, where the anisotropy is almost washed
out, and is experimentally undetectable.

\begin{figure}[t]
\centerline{\includegraphics[height=5.5cm]{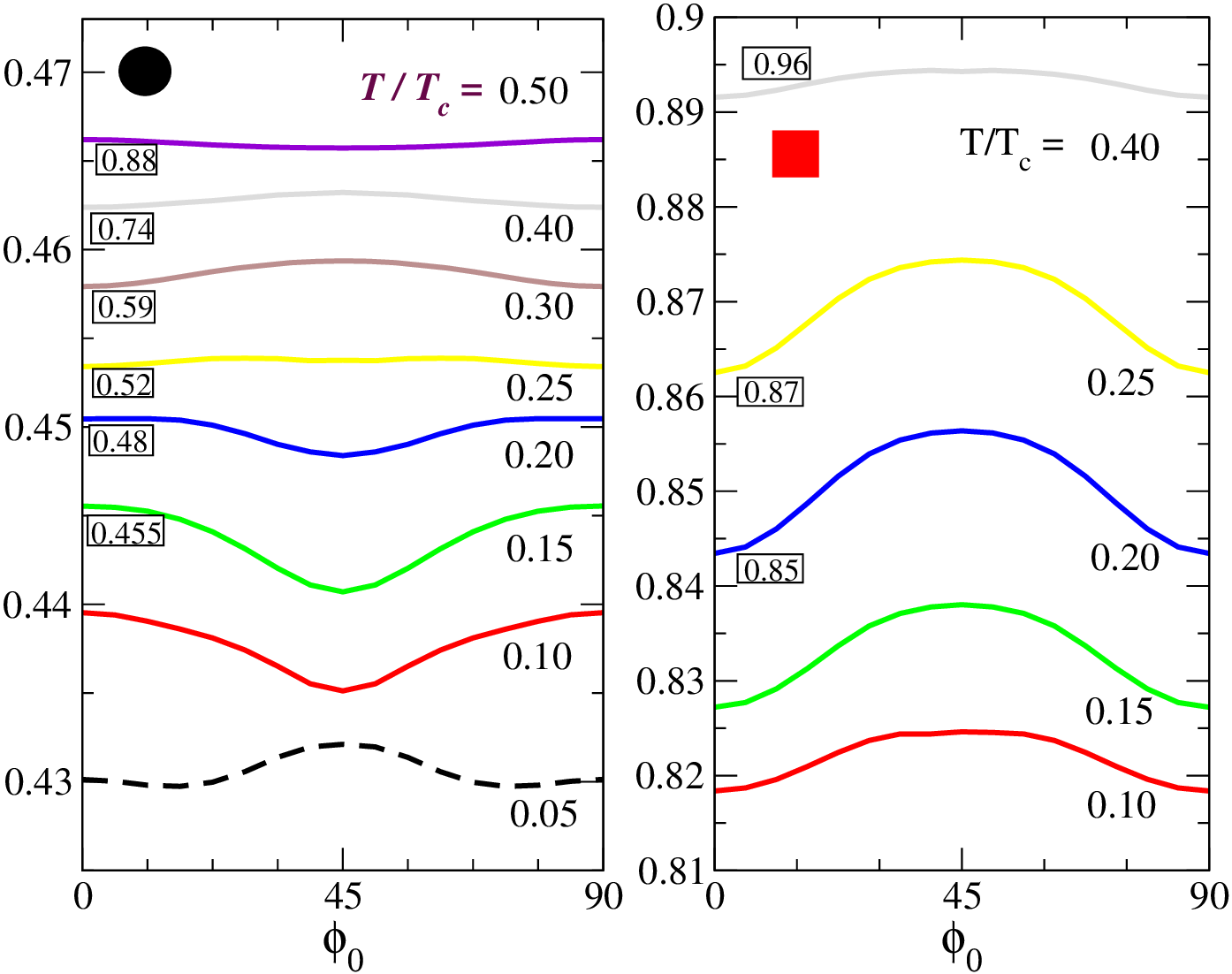}}
\caption{\label{fig:Cprof} (Color online) Heat capacity profiles
for $T$-scans at {\bf H=} and {\bf H=}. The gap node is at $\phi_0
= 45^\circ$. Left: the evolution of anisotropy with temperature
along a line of circles (low field) in Fig.\ref{fig:PDqcylC}(a).
Right: same for the high field (squares in
Fig.\ref{fig:PDqcylC}(a)). The curves that were shifted down to
appear on the same scale are shown with their original values at
$\phi_0=0$ in boxes. }
\end{figure}

Specific heat as a function of the field direction is shown in
Fig.~\ref{fig:Cprof}. The curves are computed at the $(T,H)$
points indicated in the phase diagram of Fig.~\ref{fig:PDqcylC} by
circles and squares. The left (right) panel refers to lower
(higher) field. At higher fields the gap nodes always correspond
to maxima of $C$. At low fields, however, nodes correspond to
either minima or maxima of $C$ depending on the temperature,
Fig.~\ref{fig:Cprof}(left). The lowest dashed curve in the left
panel of Fig.~\ref{fig:Cprof} appears to contradict the
semiclassical results; we show below that it corresponds to the
breakdown of the BPT approximation.  We conclude that the optimal
range of field and temperature for experimental detection of the
nodes based on the heat capacity anisotropy is at intermediate
values of $H/H_{c2}$ and $T/T_c$, where the anisotropy of $C$ is
large and the ambiguity in interpretation is small.

%%%%%%%%%%%%%%%%%%%%%%%%%%%%%%figure
%%%%%%%%%%%%%%%%%%%%%%%%%%%%%%figure

\begin{figure}[t]
\centerline{\includegraphics[height=5.5cm]{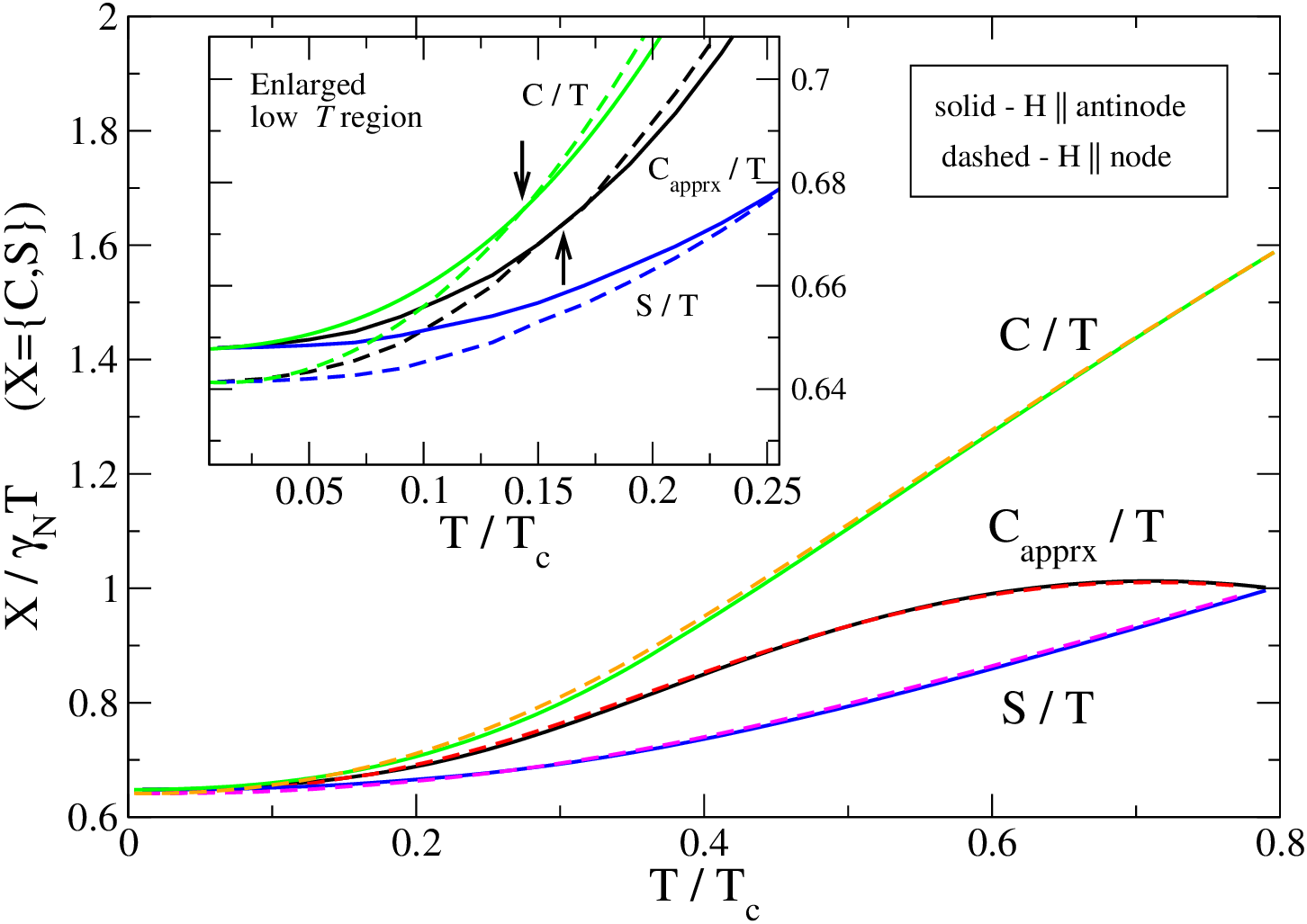}}
\caption{\label{fig:CS} (Color online) The heat capacity at
$H/H_{c2}=0.27$. Comparison of the approximate formula
Eq.~(\ref{eq:C}) with the rigorous calculation of $C/T = \partial
S/\partial T$ from numerical differentiation of the entropy. The
crossing of $C$'s for nodal and antinodal $\vH$ directions at low
temperature $T/T_c \approx 0.15$, indicated by arrows in the
inset, is not significantly affected by the approximation. }
\end{figure}

The discrepancy between $t=0.5$ profile on the left that shows a
weak minimum at the nodes and the position of the point in the
`maximum' region of the phase diagram in Fig.~\ref{fig:PDqcylC} is
due to the fact that we computed and differentiated entropy to
determine the phase diagram, but employed the approximate formula
Eq.~(\ref{eq:C}) to calculate the heat capacity anisotropy profile
(neglecting the derivative of the DOS with temperature).
Comparison of the exact and approximate formulas for the heat
capacity is shown in Fig.~\ref{fig:CS}. The lower inversion
between the minimum and the maximum of $C$ for the field along the
nodes is only slightly shifted to higher $T$ due to use of the
approximate formula (inset). The point of the high-$T$ inversion
is more sensitive to it, but, as discussed above, is not in the
regime of experimental interest.
%%%%%%%%%%%%%%%%%%%%%%%%%%%%%%figure
%%%%%%%%%%%%%%%%%%%%%%%%%%%%%%figure

At the lowest fields and temperatures in Fig.~\ref{fig:PDqcylC}
(below 0.1$H_{c2}$ and 0.07$T_c$) there appears a very small
anomalous region where the heat capacity anisotropy is inverted
compared to the semiclassical result. Our analysis shows that this
is an artefact caused by the breakdown of the BPT approximation.
Manifestations of this failure are enhanced (compared to
cylindrical FS) by the fully self-consistent calculation of the
multiple Landau channel order parameter.

\begin{figure}[t]
\centerline{\includegraphics[height=5.5cm]{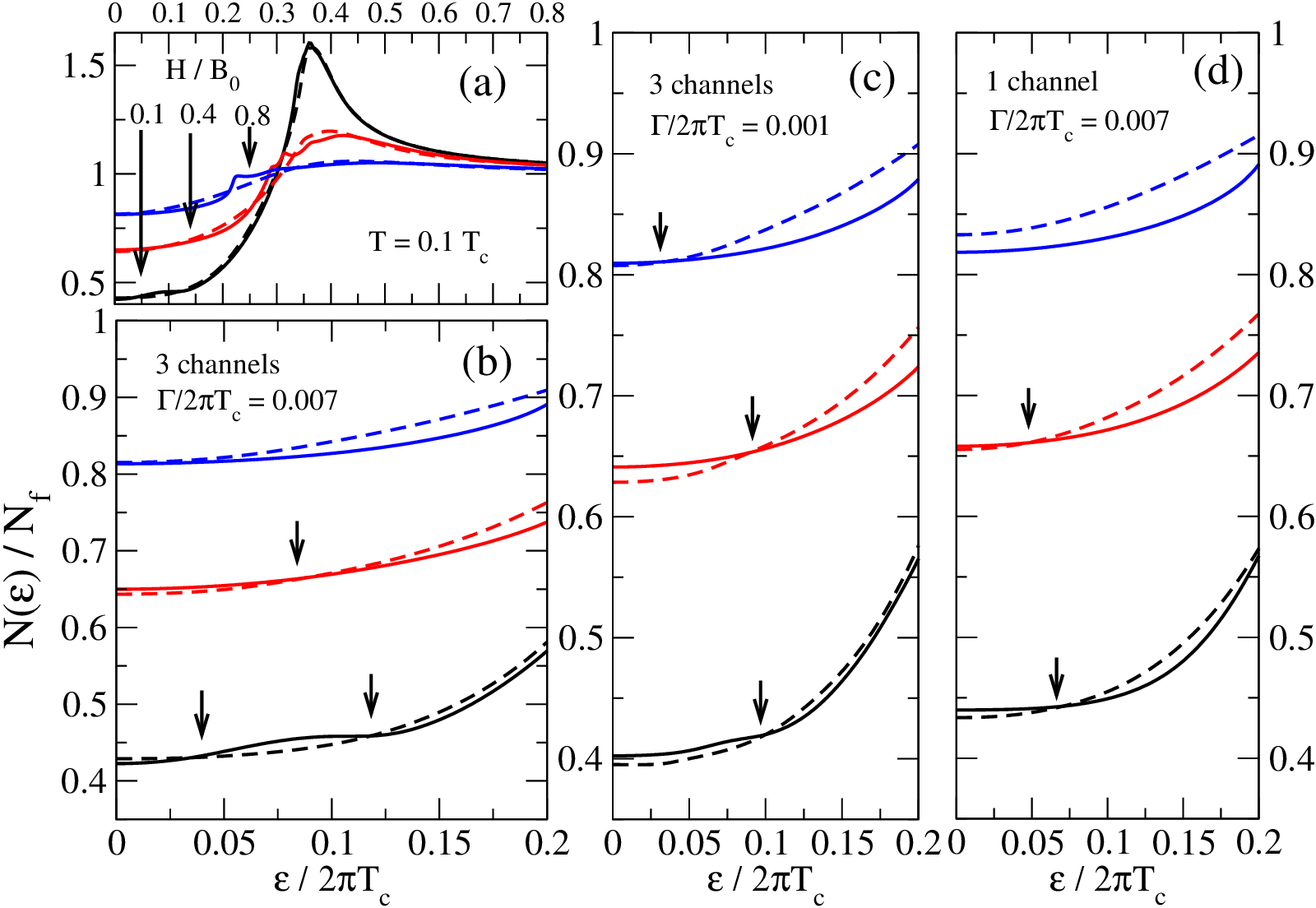}}
\caption{\label{fig:DOS} (Color online) The DOS for
quasi-cylindrical FS for fields $H/B_0=0.1, 0.4, 0.8 $
($H_{c2}=1.45 B_0 (1.27 B_0)$ for $\vH ||\,$antinode (node)).  (a)
Overall shape of the DOS for three fields, and (b) enlarged
low-energy region. The low field DOS for nodal (dashed) and
antinodal (solid) directions of $\vH$ cross several times. We
ascribe it to high derivatives of $W$-function. The oscillating
behavior decreases as system becomes cleaner (c), and disappears
for single channel (d). } \end{figure}

A necessary condition for the validity of the BPT approximation is
that the electron mean free path is much greater than the
intervortex distance, $\ell(H) = v_f \tau(H) \gg \Lambda(H)$. Only
in this case we are allowed to carry out the vortex lattice
spatial averaging before averaging over the impurity
configurations to compute the self-energy. Consequently, for
finite impurity concentration the approximation is bound to fail
at low fields. In figure \ref{fig:DOS} we present the DOS at low
field and temperature for different number of $\Delta$-channels
and the purity of the material. Notice that for the dirtier
material with more than one channel of the order parameter the DOS
oscillates at low energies when the field $\vH$ is $ || \,
antinode$, panel (b). These oscillations lead to the additional
unphysical inversion of the heat capacity anisotropy at very low
$T,H$, that is seen in the bottom left corner of the phase diagram
in Fig.~\ref{fig:PDqcylC}, and is shown by the dashed line in the
left panel of Fig.~\ref{fig:Cprof}. The same oscillations are also
present in the self-consistently calculated impurity
self-energies, which we do not show here. We find that they
decrease in a cleaner material, Fig.~\ref{fig:DOS}(c).

%%%%%%%%%%%%%%%%%%%%%%%%%%%%%%figure
%%%%%%%%%%%%%%%%%%%%%%%%%%%%%%figure

The interval of the fields where the oscillations are observed
coincides with the region where the BPT approximation is no longer
trustable. We consider the BPT breakdown at low temperature, where
the density of states is, Eq.~(\ref{eq:g-one}), $N(0,H)/N_0 =-
\Im\langle g/\pi \rangle \sim \sqrt{H/H_{c2}}$. The impurity
self-consistency in the unitary limit gives $\tau(H) = {1 \over
2\gamma} \; N(0,H)/N_0 \sim {1\over 2\gamma} \, \sqrt{H\over
H_{c2}}$. Here $\gamma=\Gamma\sin^2\delta_0$ is the normal state
scattering rate. Recalling that $\Lambda(H)/\xi_0 \sim
\sqrt{H_{c2}/H}$ and requiring $v_f \tau(H) \gg \Lambda(H)$, we
obtain a condition for the applicability of BPT: $H/H_{c2} \gg
\gamma/2\pi T_c$.
%%%%
Thus, for our impurity bandwidth $\gamma/2\pi T_c \sim 0.01$ the
BPT approximation is only applicable for fields $H/H_{c2} \gg
0.01$ and the oscillations seen in the DOS likely are a signature
of this breakdown. We checked that increasing disorder expands the
anomalous region and is consistent with this interpretation.
%%%%
For  the single-component (lowest Landau level) $\Delta$, the
numerically computed DOS does not show significant anomalous
behavior, Fig.~\ref{fig:DOS}(d), at the same impurity level as in
Fig.~\ref{fig:DOS}(b). We argue that although the breakdown of the
approximation is still there, its manifestation is less pronounced
compared with the multiple-channel order parameter. Use of higher
Landau channels for the expansion of $\Delta$ leads to the
appearance of the higher derivatives of the $W(z)$-function,
$W^{(n)}(z)$, in the Green's function, see (Eq.\ref{Eq:g-res}).
These grow very fast with $n$ at $z=0$ (approximately as $n!!$)
and are strongly oscillating as the argument is increased from
zero. This is likely the underlying reason for the oscillations in
the DOS, and therefore the ultra-low $T$-$H$ inversion is an
artefact of using the approximation beyond its region of validity.
In contrast, other inversion lines in the phase diagram correspond
to physical inversion of the measured properties.

%~~~~~~~~~~~~~~~~~~~~~~~~~~~~~~~~~~~~~~~~~~~~~~~~~~~~~~~~~~~~~~~~~~~~~~~~~
\section{\label{sec:CON} Discussion and conclusions}
%~~~~~~~~~~~~~~~~~~~~~~~~~~~~~~~~~~~~~~~~~~~~~~~~~~~~~~~~~~~~~~~~~~~~~~~~~

In this work we laid the foundations for an approach that provides
a highly flexible basis to the calculation, on equal footing, of
the transport and thermodynamic properties of unconventional
superconductors under magnetic field. The theoretical method is
based on the quasiclassical theory of superconductivity and the
Brandt-Pesch-Tewordt approximation for treatment of the vortex
state in superconductors. This approximation allows for accurate
and straightforward (analytic closed form expressions for the
Green's functions) way to describe effects of the magnetic field
in almost the entire $T$-$H$ phase diagram for clean
superconductors, with the exception of ultralow fields and
temperatures. Combined with the non-equilibrium Keldysh
formulation of the quasiclassical theory it paves a path for a
very effective computational scheme that self-consistently takes
into account multiple Landau levels of the expansion of the order
parameter and impurities, and allows calculations for arbitrary
temperature and magnitude of the field. The companion paper II
extends the method to the calculation of transport properties and
focuses on the electronic thermal conductivity.

\begin{figure}[t]
\centerline{\includegraphics[height=5.0cm]{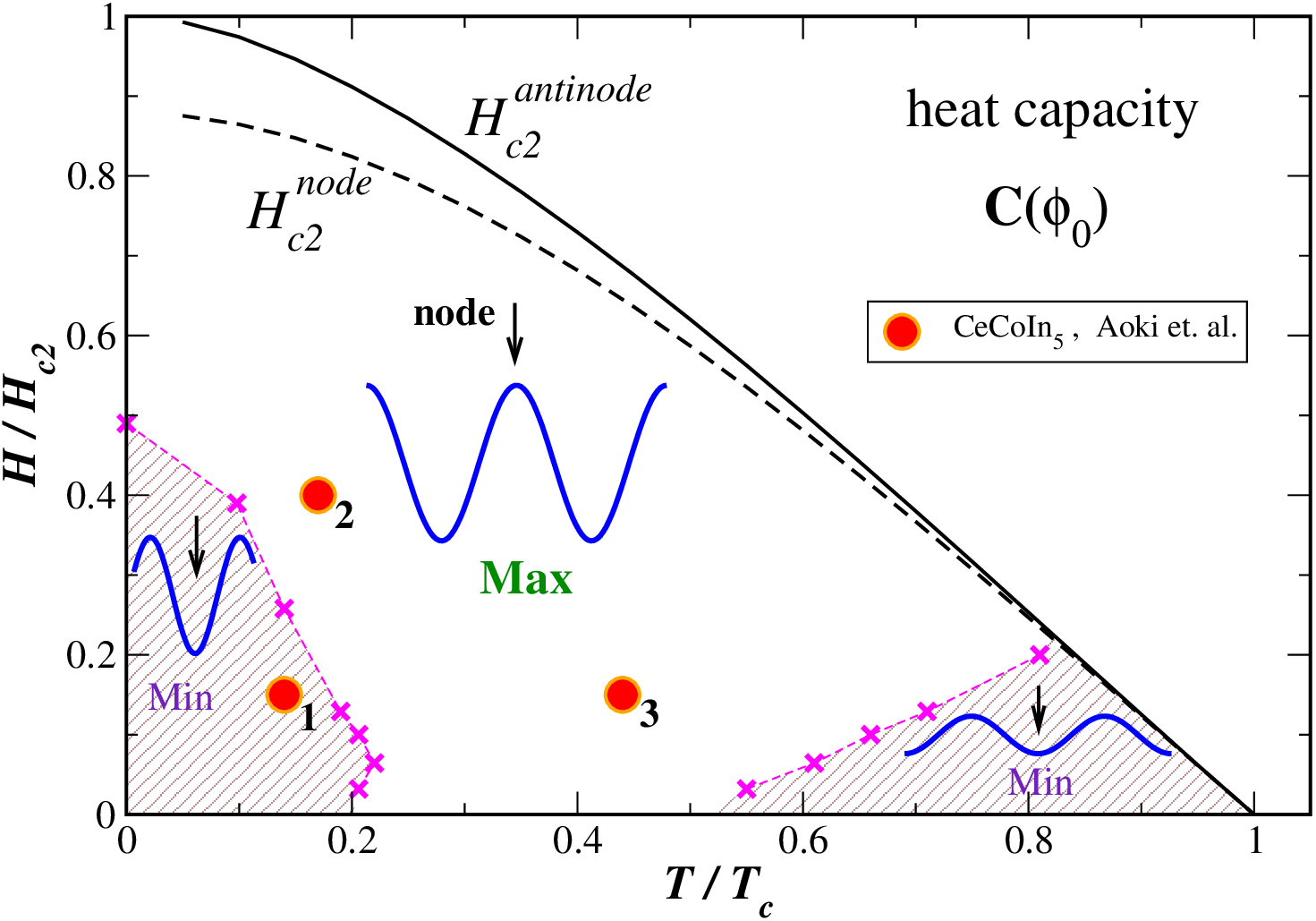}}
\caption{\label{fig:PDC} (Color online)
The anisotropy of the heat capacity in rotated magnetic field.
The $T$-$H$ phase diagram for a $d$-wave superconductor with
a corrugated cylinder Fermi surface with purely orbital depairing.
In the large part of the phase diagram the maxima of $C(\phi_0)$ function
correspond the nodal directions (indicated by arrows) on the Fermi surface.
This result is opposite to that in the Doppler region, which is confined
to $H/H_{c2}\lesssim 0.5$ and $T/T_c \lesssim 0.2$.
}
\end{figure}

Here we computed the density of states and the specific heat in
the $T$-$H$ plane for a $d$-wave superconductor with a quasi-two
dimensional Fermi surface (cylinder modulated along the symmetry
axis), and the magnetic field rotated in the basal plane. This
choice of the Fermi surface and the field orientation was
motivated by experiments on the heavy fermion CeCoIn$_5$
\cite{HAoki:2004}. We provided the first complete description of
the evolution of the anisotropy of the heat capacity due to nodes
of the superconducting gap across the $T$-$H$ phase diagram, see
Fig.~\ref{fig:PDC}.

Our main conclusion is that the anisotropic scattering of
quasiparticles due to vortices plays a crucial role in the
variation of the density of states and the specific heat as a
function of the field direction. This effect is absent in the
semiclassical (Doppler shift) approach, and becomes important
already at moderately low fields, and at finite temperatures. As
our phase diagram of Fig.~\ref{fig:PDC} shows, as a result of this
scattering, the anisotropy in the specific heat changes sign as a
function of $T$ and $H$. At low fields and temperatures the minima
in the heat capacity occur when the field is oriented along the
nodal directions, in agreement with the semiclassical (Doppler
shift) calculation. At higher $T$ and $H$ (already at $T/T_c
\gtrsim 0.2$ at low fields) the situation is reversed, and the
{\em maxima rather than minima} of the specific heat are found
when the field is along a nodal direction. Moreover, we showed
that the inversion is related to the behavior of the density of
states at finite energy, and not simply the residual DOS at the
Fermi surface.

%%%%%%%%%%%%%%%%%%%%%%%%%%%%%%figure
%%%%%%%%%%%%%%%%%%%%%%%%%%%%%%figure

While we expect that the loci of the inversion lines in the
$T$-$H$ plane weakly depend on the shape of the Fermi surface
\cite{MUdagawa:2005}, it is the existence of this inversion and
its connection to the scattering and the finite energy DOS that
emerged from our theoretical description and was not captured by
previous approaches to the problem. Our calculations serve as a
basis for the analysis of the experimental data, and we note that
the interpretation of experiments based on the low-field
expectations of minima for the field along the nodes can lead to
diametrically opposite conclusions regarding the gap symmetry,
depending on the values of the field and temperature where the
anisotropy is been measured. The results suggest that the
amplitude of the anisotropy is the greatest at intermediate
temperatures and fields, and that it is desirable not only to
measure the $C$ anisotropy at a few temperatures and fields, but
also determine its evolution over the phase diagram.

As an example, we consider the data for CeCoIn$_5$ from
Ref.~\onlinecite{HAoki:2004}, and plot in Fig.~\ref{fig:PDC} the
points where the published data were taken. The measured
$C(\phi_0)$ shows minima for the field along the [100] and [010]
directions ($\phi_0=\pi n/2$ with $n=0,1,2,3$), at all three
locations, with vanishingly small anisotropy at point 3. Points 2
and 3 are clearly in the region where maxima of $C(\phi_0)$
determine the nodes, and thus firmly point towards $d_{x^2-y^2}$
symmetry. Point 3 is also close to the inversion line, which
explains small amplitude of the oscillations.  Point 1, in
contrast, is in the ``semiclassical'' region, and therefore the
minima of $C(\phi_0)$ for the field along the crystal axes may be
more suggestive of a $d_{xy}$ symmetry. We note, however, that the
exact location of the inversion line is sensitive to the exact
shape of the Fermi surface, and changes between the calculations
restricted to the lowest Landau level for cylindrical Fermi
surface, Fig.~\ref{fig:PDcylC} and the multicomponent quasi-2D
case, Fig.~\ref{fig:PDqcylC}. We argue therefore that points 2 and
3 are more reliable indicators of the gap symmetry, and the
results are more suggestive of the $d_{x^2-y^2}$ gap. While such a
conclusion purely from the specific heat data is not foolproof, we
show in II that the $d_{x^2-y^2}$ symmetry is also supported by
the analysis of the heat transport anisotropy of
Ref.~\onlinecite{KIzawa:CeCoIn5}.

The microscopic approach, by its very nature, couples the gap
symmetry with the shape of the Fermi surface. For that reason
direct comparison of our results with other experimental data, for
example in the borocarbides YNi$_2$B$_2$C\cite{TPark:2003} and
LuNi$_2$B$_2$C\cite{TPark:2004}, is not possible. These systems
are essentially three dimensional, and the Fermi surface has no
quasi-cylindrical sheets. Moreover, it is very likely that there
is substantial gap modulation along the $z$-axis, and comparison
should be made with both point and line node models
\cite{TPark:2003,KIzawa:YNiBCkappa}. While the argument for the
change in the anisotropy due to scattering on the vortices is
quite general, the position of the anisotropy inversion lines in
the phase diagram (if any) is undoubtedly different from that
found for the quasi-2D system, and such differences are known to
occur in the zero-energy DOS ~\cite{MUdagawa:2005}. Therefore we
will consider the nodal structures of these systems separately in
near future.

To reiterate, the approach described in this work presents a
powerful tool to study the gap symmetry in the unconventional
superconductors taking into account their realistic Fermi
surfaces. Our results serve as a basis for interpretation of
experimental data, pointing towards a resolution of the
discrepancy between the results of the specific heat and thermal
conductivity measurements in CeCoIn$_5$, which is also addressed
in II.  The method developed here can be easily generalized to
include other Fermi surfaces, paramagnetic effects, and other
aspects of real materials, the discussion of which we defer to
future publications.

%~~~~~~~~~~~~~~~~~~~~~~~~~~~~~~~~~~~~~~~~~~~~~~~~~~~~~~~~~~~~~~~~~~~~~~~~~
\section{Acknowledgements}
%~~~~~~~~~~~~~~~~~~~~~~~~~~~~~~~~~~~~~~~~~~~~~~~~~~~~~~~~~~~~~~~~~~~~~~~~~

This work was partly done at KITP with support from NSF Grant
PHY99-07949;  and was also supported by the Board of Regents of
Louisiana. We thank D.~A.~Browne, C.~Capan, P.~J.~Hirschfeld,
Y.~Matsuda, and T.~Sakakibara for discussions, and one of us
(I.~V) is very indebted to A. Houghton for encouragement at the
early stages of this work.

%~~~~~~~~~~~~~~~~~~~~~~~~~~~~~~~~~~~~~~~~~~~~~~~~~~~~~~~~~~~~~~~~~~~~~~~~~
%~~~~~~~~~~~~~~~~~~~~~~~~~~~~~~~~~~~~~~~~~~~~~~~~~~~~~~~~~~~~~~~~~~~~~~~~~
\appendix
\section{\label{app:GL} Choice of operators for anisotropic Fermi surface}

The raising and lowering operators for the eigenfunction expansion
of the order parameter, $a^\dag, a$ and the corresponding ladder
states can be introduced in several different ways. We want to
define them in the manner that facilitates the efficient
computations. This issue becomes important for anisotropic Fermi
surfaces and arbitrary direction of the field. Anisotropy of the
FS is directly translated into the shape of a single vortex and we
can choose the orthogonal states such that they approximate this
shape well already at the lowest order truncation of the expansion
of $\Delta(\bm R)$.

We consider an axisymmetric FS in cylindrical coordinates
$(r,\phi,z_0)$. The energy has the form $2m\vare = p_r^2 +
f(p_{z_0})$, with an arbitrary function $f(p_{z_0})$ of the
$p_{z_0}$ momentum. The vortex state near the critical temperature
$T_c$ is determined from the linearized GL equations,
%For the clean case in the absence of the field,
    %\bea
    %\Delta(\vR) \ln {T\over T_c} = T \sum_{\vare_m} \int d\hat{\vp} \cY^2(\hat{\vp}) \times
    %\nonumber \\
    %{ -\pi \over 4 |\vare_m|^3}
    %\left\{
    %(i\vv_f(\hat{\vp}) \grad)^2 \Delta(\vR) + \Delta^3(\vR)  \cY^2(\hat{\vp})
    %\right\}
    %\eea
%Adding the field ($\grad \to \grad -i 2|e|\vA /c$) and looking at the
%region near $H_{c2}$, we left with equation
    \bea
    -K_{ij} \tilde{\grad}_i \tilde{\grad}_j \Delta(\vR) + \left({T\over T_c} -1 \right) \Delta(\vR) = 0
    \\
    K_{ij} = T_c \sum_{\vare_m} { \pi \over 4 |\vare_m|^3}
    \langle \cY^2(\hat{\vp}) v_{f,i}(\hat{\vp}) v_{f,j}(\hat{\vp}) \rangle_\sm{FS}
    \\
    \tilde{\grad}= \grad - i{2e \over c} \vA(\vR)
    \eea
In these equations the coordinates $(x,y,z)$ are chosen so that
the field is along the $z$-axis, $\vB = B \hat{\vz}$, and we take
$\vA = (0, B x, 0)$. The form of the $K_{ij}$ tensor depends on
the shape of the Fermi surface, the pairing state and orientation
of the magnetic field. If the rotational symmetry axis is $z_0$,
the velocity of quasiparticles is $\vv_f(\phi, p_{z_0}) =
(v_r(p_{z_0}) \cos\phi, v_r(p_{z_0}) \sin\phi, v_{z_0}(p_{z_0}))$.
If $\vB$ is along one of the FS symmetry axes, $x_0$, $y_0$ or
$z_0$, $K_{ij}$ is diagonal for $d$-wave pairing with $\cY =
\sqrt{2}\cos2(\phi-\phi_0)$. We have, $K_{x_0 x_0} = K_{y_0 y_0} =
K_0 v_{0\perp}^2$, $K_{z_0 z_0} = K_0 v_{0||}^2$, where
$K_0=7\zeta(3)/8(2\pi T_c)^2$, and
    \bea
    v_{0\perp}^2 &=&  2 \, \langle \cY^2(\hvp) v^2_{x_0(y_0)}(p_{z_0}) \rangle_\sm{FS} \,,
    \\
    v_{0\parallel}^2 &=& 2 \, \langle \cY^2(\hvp) v^2_{z_0}(p_{z_0}) \rangle_\sm{FS} \,.
    \eea
We apply the magnetic field at a tilt angle $\theta_H$ from $z_0$
direction towards $x_0$ axis. A coordinate system associated with
$\vB$ is chosen as follows, $\hat{z}$ is along $\vB$, $\hat{y} =
\hat{y}_0$ and $\hat{x}$ lies in $(x_0,z_0)$-plane and
perpendicular to $\hat{z}$. Projections of the Fermi velocity at
different points of the FS on these new coordinate axes,
$v_{f,x}(p_{z_0},\phi) = v_{f,x_0}(p_{z_0},\phi) \cos\theta_H -
v_{f,z_0}(p_{z_0},\phi) \sin\theta_H$, $v_{f,y}(p_{z_0},\phi) =
v_{f,y_0}(p_{z_0},\phi)$, $v_{f,z}(p_{z_0},\phi) =
v_{f,z_0}(p_{z_0},\phi) \cos\theta_H + v_{f,x_0}(p_{z_0},\phi)
\sin\theta_H$. In $(x,y,z)$-coordinates the tensor $K_{ij}$ is not
diagonal anymore, $K_{yy} = K_{y_0 y_0}$, $K_{xx} = K_{x_0 x_0}
\cos^2 \theta_H + K_{z_0 z_0} \sin^2 \theta_H$, $K_{zz} = K_{x_0
x_0} \sin^2 \theta_H + K_{z_0 z_0} \cos^2 \theta_H$, $K_{xz} =
(K_{x_0 x_0} - K_{z_0 z_0}) \sin\theta_H \cos\theta_H$, and for
the choice of the operator, $\grad - i2e/c \, \vA = (\grad_x,
\grad_y - i 2e/c \, B x, \grad_z )$, the GL equation is
    \bea
    -K_{xx} \grad^2_x \; \Delta -K_{yy} \left(\grad_y - i{2eB\over c} x \right)^2 \; \Delta
    \nonumber \\
    -2 K_{xz} \grad_x \grad_z \; \Delta  + \left({T\over T_c} -1 \right) \; \Delta = 0 \,.
    \label{eq:GLxy}
    \eea
%%%%%%%%%%%%%%%%%%%%%%%%%%%%%%%figure
\begin{figure}[t]
\centerline{\includegraphics[height=5.5cm]{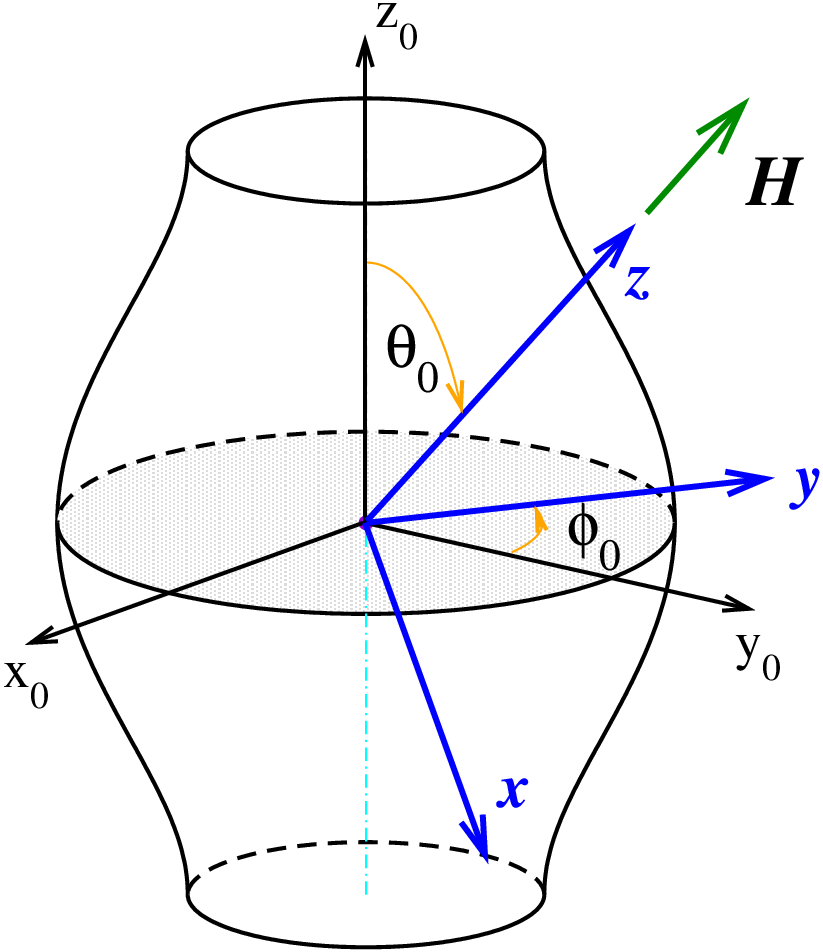}}
\caption{\label{fig:FSshape} (Color online)
Quasi-cylindrical Fermi surface considered in this paper.
Direction of $\vH$ defines $(xyz)$ coordinates with $\hat{\vz} || \vH$.
To go from $(x_0 y_0 z_0)$ coordiantes, associated with the Fermi surface, to
$(xyz)$ coordinates, associated with the field, we perform first
rotation by $\phi_0$ around $\hat{\vz}_0$, and then rotation
by $\theta_0$ around $\hat{\vy}$.
}
\end{figure}
%%%%%%%%%%%%%%%%%%%%%%%%%%%%%%%figure
It is easy to check by setting $\Delta = \Delta(x,y) \exp(ik_z z)$
that the highest critical field still corresponds to $k_z=0$, and
we put $\grad_z\Delta=0$ below. We rescale the coordinates
$x'=x/\sqrt{S_f}$ and $y'= y \sqrt{S_f}$, and choose the scaling
factor $S_f$ such that $K_{xx}/S_f = K_{yy} S_f$. Thus,
    \be
    S_f^2 = {K_{xx}\over K_{yy}} =
    \cos^2 \theta_H + \frac{v_{0\|}^2}{v_{0\perp}^2} \sin^2 \theta_H  \,.
    \ee
After introducing creation and annihilation operators ($\Lambda^2 = c/2|e|B$ and $e<0$),
    \bea
    a = {\Lambda\over \sqrt{2}} \left( -\grad_{x'} + i (\grad_{y'} + i{x'\over \Lambda^2}) \right) \,,
    \\
    a^\dag = {\Lambda\over \sqrt{2}} \left( \grad_{x'} + i (\grad_{y'} + i{x'\over \Lambda^2}) \right) \,,
    \eea
%These operators obey commutation relations,
    %\be
    %[a,a^\dag] = 1 \qquad,\qquad [a,a] = 0 \,,
    %\ee
%and can be used to construct our Hilbert space.
equation (\ref{eq:GLxy}) becomes
    \be
    \left( a^\dag a + {1\over2} \right) \Delta(x,y) = {\Lambda^2\over2} {S_f\over K_{xx}}
    \left(1- {T\over T_c} \right) \; \Delta(x,y) \,.
    \ee
Then for {\it any} axisymmetric FS we obtain the well-known result
of the anisotropic mass model for the upper critical field,
determined by the ratio of the Fermi velocities for the two
directions,
    \be
    B_{c2}(\theta_H,T) = {const \cdot (1-T/T_c) \over
    \sqrt{\cos^2\theta_H + {v_{0||}^2\over v_{0\perp}^2}\sin^2\theta_H} }
    \,.
    \ee
We also find a set of eigenfunctions,
    \be
    \Delta^{(n)}(x,y) = \sum_{k_y} C_{k_y}^{(n)} {e^{ik_y\sqrt{S_f} y} \over \sqrt[4]{S_f \Lambda^2}}
    \Phi_n\left( {x-\Lambda^2 \sqrt{S_f} k_y\over \Lambda \sqrt{S_f}} \right) \,.
    \label{eq:staten}
    \ee
In terms of the operators ${a, a^\dag}$ the gradient term in the
Eilenberger equation has the form
    \bea
    \lefteqn{ \vv_f(\hat{\vp})\,(\gradR -   i {2e\over c} \vA(\vR) ) = } &&
    \nonumber \\
    && {|\tilde{v}_f^\perp| \over \sqrt{2} \Lambda}
    [-\tilde{v}_{+}(\hat{\vp}) \, a + \tilde{v}_{-}(\hat{\vp}) \, a^\dag] \,.
    \label{eq:gradLadder}
    \eea
Here we   rescaled the Fermi velocity in the $xy$-plane
    \bea
    \tilde{v}_f(\hat{\vp})_x &=& v_f(\hat{\vp})_x /\sqrt{S_f} \, ,\\
    \tilde{v}_f(\hat{\vp})_y &=& v_f(\hat{\vp})_y \sqrt{S_f}\, ,
    \eea
with
    \be
    |\tilde{v}_f^\perp (\hat{\vp})| =
    \sqrt{\tilde{v}_f(\hat{\vp})_x^2 + \tilde{v}_f(\hat{\vp})_y^2}\,,
    \ee
and
    \be
    \tilde{v}_\pm (\hat{\vp})=
    \frac{\tilde{v}_f(\hat{\vp})_x \pm i \tilde{v}_f(\hat{\vp})_y}
         {|\tilde{v}_f^\perp|} \,.
    \ee

\section{\label{app:F} Closed form solution for the Green's function}

To solve the semiclassical equations we use
Eq.~(\ref{eq:gradLadder}) to cast the operator $\hat{O}_f$ from
Eq.~(\ref{eq:O}) in an integral form,
    \bea
    \hat{O}_f &=& [-2 i \tilde{\vare} + {|\tilde{v}^\perp_f| \over \sqrt{2} \Lambda}
    (\tilde{v}_- a^\dag -\tilde{v}_+ a)]^{-1}
    \nonumber \\
    &=& \int\limits_0^\infty d t_1 \;
    e^{ - [-2 i \tilde{\vare} + {|\tilde{v}^\perp_f| \over \sqrt{2} \Lambda}
    (\tilde{v}_- a^\dag -\tilde{v}_+ a) ] t_1}
    \nonumber \\
    &=& \int\limits_0^\infty d t_1 \;
    e^{ 2 i \tilde{\vare} t_1 - {w^2 } t_1^2/2 }
    e^{ - w t_1 \, \tilde{v}_- a^\dag }  e^{ w t_1 \, \tilde{v}_+ a } \,,
    \eea
where we introduced the magnetic field energy
    \be
    w = {|\tilde{v}^\perp_f| \over \sqrt{2} \Lambda}\,
    \ee
and used the operator identity $\exp(A+B) = \exp(A) \exp(B)
\exp(-C/2)$, if $[A,B]=C$ is a $c$-number. In all integrals we
also keep in mind that $\Im \tilde{\vare} >0$ for {\it retarded}
functions, so that the convergence is ensured.

In this formulation it is convenient to work with bra- and ket-functions for different
vortex states, which correspond in $\vR$-representation to states (\ref{eq:staten}).
We present decomposition of $\tilde{\Delta}$ as
    \be
    \widetilde{\Delta} = \sum_n \widetilde{\Delta}_n \ket{n} \,,
    \qquad
    \widetilde{\ul\Delta} = \sum_n \widetilde{\ul\Delta}_n \bra{n} \,,
    \ee
with operator $\hat{O}_f$ acting to the right,
    \bea
    f &=& (2ig)\,  \hat{O}_f \widetilde{\Delta} =
    \\
    &=& (2ig) \int\limits_0^\infty d t_1 \, e^{ 2 i \tilde{\vare} t_1 - {w^2} t_1^2 / 2 }
    e^{ - w t_1 \, \tilde{v}_- a^\dag }  e^{ w t_1 \, \tilde{v}_+ a }
    \widetilde{\Delta} \,,
    \nonumber
    \eea
and operator $\hat{O}_{\ul{f}}^\dag$ acting to the left,
    \be
    \ul{f} = (2ig)\,  \widetilde{\ul\Delta} \, \hat{O}_{\ul{f}}^\dag \,.
    \ee
We rewrite the operator $\hat{O}_{\ul{f}}$  as
    \bea
    \hat{O}_{\ul{f}} &=& [2 i \tilde{\vare}^* - w (\tilde{v}_- a^\dag -\tilde{v}_+ a)]^{-1}
    = \hat{O}^\dag_f =
    \\
    &=& \int\limits_0^\infty d t_2 \;
    e^{ -2 i \tilde{\vare}^* t_2 - {w^2 } t_2^2/2 }
    e^{ w t_2 \, \tilde{v}_- a^\dag }  e^{ -w t_2 \, \tilde{v}_+ a } \,.
    \nonumber
    \eea
so that the spatial average of the off-diagonal functions is
    \bea
    \overline{\ul{f} \, f} &=&
    (2i g )^2 \int\limits_0^\infty d t_1 \, \int\limits_0^\infty d t_2 \;
    e^{ 2 i \tilde{\vare} (t_1+t_2) - {w^2 } (t_1+t_2)^2 /2 } \times
    \nonumber \\
    &\times& \widetilde{\ul\Delta} \left(
    e^{ - w (t_1+t_2) \, \tilde{v}_- a^\dag }  e^{ w (t_1+t_2) \, \tilde{v}_+ a }
    \right) \widetilde{\Delta} \,,
    \eea
where we make sure that bra-vectors stay on the left of ket-vectors.
Here we again used operator-in-exponent rule to commute exponents.
After an appropriate variable susbstitution,
    \be
    \overline{\ul{f} \, f} =
    (2i g )^2 \int\limits_0^\infty d t \; t \,
    e^{ 2 i \tilde{\vare} t - {w^2 } t^2/2 }
    \widetilde{\ul\Delta} \left(
    e^{ - w t \, \tilde{v}_- a^\dag }  e^{ w t \, \tilde{v}_+ a }
    \right) \widetilde{\Delta} \,.
    \ee
This form is very convenient if we intend to keep several Landau
channels in the expansion of $\Delta$. If the highest Landau level
used is  $N$, the series expansion for $e^{ w t \, \tilde{v}_+ a
}$ contains only $N+1$ terms; and to calculate the spatial average
$\overline{\ul{f} \, f}$ we need to compute only a finite number,
($2N+1$), of $W^{(n)}$-functions, since
    \bea
    \lefteqn{
    \int\limits_0^\infty d t \; t \, (w t)^n \, e^{ 2 i \tilde{\vare} t - {w^2 } t^2/2 }
    =}&& \nonumber \\
    && = {1\over 2 w^2} (-i\sqrt{\pi}) \left(-{i\over \sqrt{2}} \right)^n
    W^{(n+1)}\left( {\sqrt{2} \tilde{\vare} \over w} \right) \,.
    \label{eq:Wn}
    \eea

Solution for $f$ is written as,
    \be
    f(\vR, \hat{\vp}; \vare) = \sum_m f_m(\hvp, \vare) \braket{\vR}{m} \,,
    \ee
with the amplitudes
    \be
    f_m(\hvp, \vare) = ig \, \sum_n (-\tilde{v}_-)^{m-n}(\hvp) \,
        \cD_{m,n}(\vare, |\hvp|) \widetilde{\Delta}_n(\hvp; \vare)
        \,.
    \label{eq:oscf}
    \ee
Here
    \be
    \cD_{m,n}(\vare, |\hvp|) = {2\sqrt{\pi} \Lambda \over |\tilde{v}^\perp_f|}
    \sum_{n_2}^{m} (-1)^{n_1}
    D_n^{n_1, n_2} \left({2\tilde{\vare}\Lambda\over |\tilde{v}^\perp_f|} \right) \,,
    \ee
    \be
    D^{n_1,n_2}_{n}(z) =
    \left({-i\over\sqrt{2}}\right)^{n_1+n_2}
    {\sqrt{n!} \sqrt{(n-n_1+n_2)!} \over (n-n_1)! n_1! n_2!}
    W^{(n_1+n_2)}(z).
    \ee
The sum starts from $n_2 = max(0,m-n)$ and, in each term,
$n_1=n-m+n_2$. This sum can be cast in a more symmetric form with
respect to the indices $m,n$, which we present in the main text in
Eq.(\ref{eq:Dmn}).

%Solution is known to be
    %\bea
    %&&f = 2 i g \sum_{n=0}^\infty \Delta_n \cY(\hat{\vp}) \left( \frac{\sqrt{\pi} \Lambda}{|v_f^\perp|} \right)
    %\times \\
    %&&\left\{
    %\sum_{n_1=0}^n \sum_{n_2=0}^\infty \, (-1)^{n_1} \, D^{n_1,n_2}_{n} (u_m) \, v_-^{n_1} v_+^{n_2}
    %\; \ket{n - n_1 +n_2}
    %\right\} \,, \nonumber
    %\label{eq:oscf}
    %\eea
We limit ourselves to superconductors with inversion symmetry.
Then the singlet and triplet order parameters transform under
inversion as follows,
    \bea
    \cP \Delta(\vR,\hvp) = \Delta(-\vR, -\hvp) = \Delta(\vR, -\hvp) = \Delta(\vR, \hvp) \,,
    \nonumber \\
    \cP \vDelta(\vR,\hvp) = \vDelta(-\vR, -\hvp) = \vDelta(\vR, -\hvp) = -\vDelta(\vR, \hvp) \,,
    \nonumber \eea
where we assumed that the order parameter is an even function of
the spatial coordinates $\vR$. This assumption is justified by the
analysis of the behavior of the off-diagonal functions
$f_m(\hvp)$. The expansion of the anomalous propagators in the
Landau level basis contains all components $\langle {\bf
R}|m\rangle$, however, even and odd coefficients have different
parity under inversion ${\bf p}\rightarrow -{\bf p}$,
    \bea
    f^{s}_m(-\hvp) = ig \, \sum_n \tilde{v}_-^{m-n}(\hvp) \, \cD_{m,n}(|\hvp|) \widetilde{\Delta}_n(\hvp) \,,
    \nonumber \\
    f^{t}_m(-\hvp) = - ig \, \sum_n \tilde{v}_-^{m-n}(\hvp) \, \cD_{m,n}(|\hvp|) \vDelta_n(\hvp)
    \,.
    \nonumber
    \eea
As a result, it is easy to show that for both singlet and triplet
order parameters, the even and odd coefficients $\Delta_n$ are
decoupled since no mixed term survives averaging over the Fermi
surface,
    \be
    \int \dangle{p} \cY_{s,t}(\hvp) f^{s,t}(\hvp).
    \ee

Note also that, in zero field for superconductors with basis
functions $\langle \cY(\hvp) \rangle_\sm{FS}=0$, the off-diagonal
impurity self-energy vanishes since $\langle f(\hvp)
\rangle_\sm{FS}=0$. Under magnetic field, however, the direction
$\hvp$ is inequivalent to the perpendicular to it direction,
$\hvp_\perp$, and $f(\hvp)$ is not simply proportional to
$\cY(\hvp)$. Hence for the field in the plane and  $d$-wave gap
$\langle f(\hvp) \rangle_\sm{FS}\neq 0$, and there is a
contribution to the off-diagonal self-energies from impurities.
This integral still vanishes for $p$-wave order parameters,
since in magnetic field the directions $\hvp$ and $-\hvp$ may
remain equivalent and so the symmetry $f(-\hvp) = -f(\hvp)$ is
still valid.

%~~~~~~~~~~~~~~~~~~~~~~~~~~~~~~~~~~~~~~~~~~~~~~~~~~~~~~~~~~~~~~~~~~~~~~~~~~~~~
%~~~~~~~~~~~~~~~~~~~~~~~~~~~~~~~~~~~~~~~~~~~~~~~~~~~~~~~~~~~~~~~~~~~~~~~~~~~~~
%\bibliographystyle{apsrev}
%\bibliography{/home/anton/script/bibliography/CM,/home/anton/script/bibliography/QFS,BPT2}
%\bibliography{BPT}

%%%%%%%%%%%%%%%%%%%%%%%%%%%%%%%%%%%%%%%%%%%%%%%%%%%%%%%%%%%%

%%%%%%%%%%%%%%%%%%%%%%%%%%%%%%%%%%%%%%%%%%%%%%%%%
\end{document}